\documentstyle[psfig]{espart}

\begin{document}

\begin{frontmatter}

\title{Modern nucleon-nucleon potentials and 
       symmetry energy in infinite matter}
 
\author[oslo]{L.\ Engvik}, 
\author[kbh]{M.\ Hjorth-Jensen}, 
\author[idaho]{R.\ Machleidt},  
\author[tuebingen]{H.\ M\"{u}ther},  
\author[barcelona]{A.\ Polls}
\address[oslo]{Department of Physics, University of Oslo, N-0316 Oslo, Norway}
\address[kbh]{Nordita, Blegdamsvej 17, DK-2100 K\o benhavn \O, Denmark}
\address[idaho]{Department of Physics, University of Idaho, Moscow, 
         ID 83844, U.S.A.}
\address[tuebingen]
         {Institut f\"ur Theoretische Physik, Universit\"at T\"ubingen,
         D-72076 T\"ubingen, Germany}
\address[barcelona]{Departament d'Estructura i Costituents de la Mat\`eria,
         Universitat de Barcelona, E-08028 Barcelona, Spain}

\maketitle

\begin{abstract}

We study the symmetry energy in infinite nuclear
matter employing a non-relativistic Brueckner-Hartree-Fock approach
and using various new nucleon-nucleon (NN) potentials, 
which fit np and pp scattering
data very accurately. The potential models we employ
are the recent
versions of the Nijmegen group, Nijm-I, Nijm-II and Reid93, 
the Argonne $V_{18}$ potential and the CD-Bonn potential.
All these potentials yield a symmetry energy which increases with density,
resolving a discrepancy that existed for older
NN potentials. The origin of remaining differences is
discussed.

\end{abstract}

\end{frontmatter}

\section{Introduction}

The equation of state (EOS) for neutron star matter and infinite
nuclear matter has been intensively studied for many years.
A correct description of the EOS would have  far reaching consequences
for topics ranging  from the cooling of neutron stars \cite{pr95,prakash94} 
to the physics of heavy ion collisions \cite{bkr97}.
Furthermore, recent radioactive ion beam experiments
\cite{tanihata95,hjj95} have provided new information on the structure
of unstable nuclei far from equilibrium. The latter may open the 
possibility of extracting information on the EOS for asymmetric 
matter and the density dependence of the nuclear symmetry energy.
The symmetry energy ${\cal S}(n)$ is defined as the difference between
the energy per particle in pure neutron matter ${\cal E}(n,\chi_p=0)$ and 
the binding energy per particle in symmetric nuclear 
matter ${\cal E}(n,\chi_p=1/2)$, where $n$ is the total baryon density
in units of fm$^{-3}$ and $\chi_p=n_p/n$ is the proton fraction. 

Knowledge of the symmetry energy at high densities is of particular importance
in nuclear astrophysics where the density dependence of 
${\cal S}(n)$ is crucial for understanding 
e.g., the cooling mechanisms in a neutron star \cite{prakash94}.
The symmetry energy determines 
the proton fraction in $\beta$-stable matter. 
The proton fraction is in turn important for the cooling history of
a neutron star. At a certain critical value, typically $\chi_p \approx 0.15$,
the so-called direct URCA processes for neutrino emissions 
are allowed \cite{prakash94}. Further, neutrino emissivity 
from the so-called modified URCA processes do also
depend on the given proton fractions, see e.g., Ref.\ \cite{fm79}. 

Many calculations of the symmetry energy and the EOS have been performed using
different methods to solve the non-relativistic many-body problem of infinite
nuclear matter and employing various realistic models of the 
nucleon-nucleon (NN) interaction,
which were all adjusted to describe NN scattering phase shifts. All these
calculations yield similar results for the symmetry energy around the 
saturation density of nuclear matter and are in reasonable 
agreement with the empirical estimate \cite{pr95}. The predictions for the
symmetry energies at high densities were quite different. The variational
calculations of Wiringa et al.\ \cite{wff88} using the Argonne
$V_{14}$ potential \cite{wsa84}, predict e.g., a symmetry energy which 
varies only very little for densities $n >$0.3 fm$^{-3}$, while
Brueckner-Hartree-Fock (BHF) calculations using One-Boson-Exchange potentials
predict  symmetry energies which increase with density also for larger
densities \cite{ehobo96,bbb96}. 

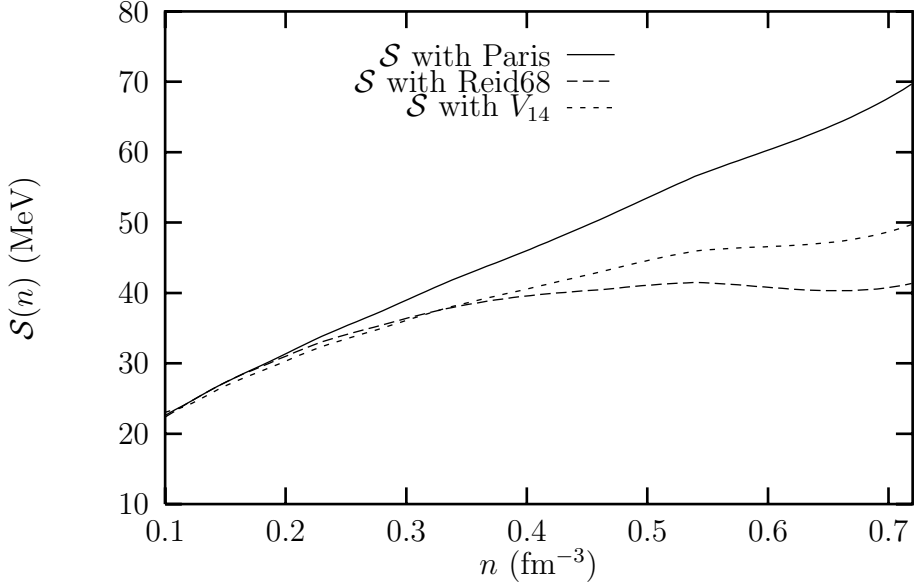
\begin{figure}[htbp]
\setlength{\unitlength}{0.1bp}
\special{!
/gnudict 40 dict def
gnudict begin
/Color false def
/Solid false def
/gnulinewidth 5.000 def
/vshift -33 def
/dl {10 mul} def
/hpt 31.5 def
/vpt 31.5 def
/M {moveto} bind def
/L {lineto} bind def
/R {rmoveto} bind def
/V {rlineto} bind def
/vpt2 vpt 2 mul def
/hpt2 hpt 2 mul def
/Lshow { currentpoint stroke M
  0 vshift R show } def
/Rshow { currentpoint stroke M
  dup stringwidth pop neg vshift R show } def
/Cshow { currentpoint stroke M
  dup stringwidth pop -2 div vshift R show } def
/DL { Color {setrgbcolor Solid {pop []} if 0 setdash }
 {pop pop pop Solid {pop []} if 0 setdash} ifelse } def
/BL { stroke gnulinewidth 2 mul setlinewidth } def
/AL { stroke gnulinewidth 2 div setlinewidth } def
/PL { stroke gnulinewidth setlinewidth } def
/LTb { BL [] 0 0 0 DL } def
/LTa { AL [1 dl 2 dl] 0 setdash 0 0 0 setrgbcolor } def
/LT0 { PL [] 0 1 0 DL } def
/LT1 { PL [4 dl 2 dl] 0 0 1 DL } def
/LT2 { PL [2 dl 3 dl] 1 0 0 DL } def
/LT3 { PL [1 dl 1.5 dl] 1 0 1 DL } def
/LT4 { PL [5 dl 2 dl 1 dl 2 dl] 0 1 1 DL } def
/LT5 { PL [4 dl 3 dl 1 dl 3 dl] 1 1 0 DL } def
/LT6 { PL [2 dl 2 dl 2 dl 4 dl] 0 0 0 DL } def
/LT7 { PL [2 dl 2 dl 2 dl 2 dl 2 dl 4 dl] 1 0.3 0 DL } def
/LT8 { PL [2 dl 2 dl 2 dl 2 dl 2 dl 2 dl 2 dl 4 dl] 0.5 0.5 0.5 DL } def
/P { stroke [] 0 setdash
  currentlinewidth 2 div sub M
  0 currentlinewidth V stroke } def
/D { stroke [] 0 setdash 2 copy vpt add M
  hpt neg vpt neg V hpt vpt neg V
  hpt vpt V hpt neg vpt V closepath stroke
  P } def
/A { stroke [] 0 setdash vpt sub M 0 vpt2 V
  currentpoint stroke M
  hpt neg vpt neg R hpt2 0 V stroke
  } def
/B { stroke [] 0 setdash 2 copy exch hpt sub exch vpt add M
  0 vpt2 neg V hpt2 0 V 0 vpt2 V
  hpt2 neg 0 V closepath stroke
  P } def
/C { stroke [] 0 setdash exch hpt sub exch vpt add M
  hpt2 vpt2 neg V currentpoint stroke M
  hpt2 neg 0 R hpt2 vpt2 V stroke } def
/T { stroke [] 0 setdash 2 copy vpt 1.12 mul add M
  hpt neg vpt -1.62 mul V
  hpt 2 mul 0 V
  hpt neg vpt 1.62 mul V closepath stroke
  P  } def
/S { 2 copy A C} def
end
}
\begin{picture}(3600,2160)(0,0)
\special{"
gnudict begin
gsave
50 50 translate
0.100 0.100 scale
0 setgray
/Helvetica findfont 100 scalefont setfont
newpath
-500.000000 -500.000000 translate
LTa
LTb
600 251 M
63 0 V
2754 0 R
-63 0 V
600 516 M
63 0 V
2754 0 R
-63 0 V
600 782 M
63 0 V
2754 0 R
-63 0 V
600 1047 M
63 0 V
2754 0 R
-63 0 V
600 1313 M
63 0 V
2754 0 R
-63 0 V
600 1578 M
63 0 V
2754 0 R
-63 0 V
600 1844 M
63 0 V
2754 0 R
-63 0 V
600 2109 M
63 0 V
2754 0 R
-63 0 V
600 251 M
0 63 V
0 1795 R
0 -63 V
1054 251 M
0 63 V
0 1795 R
0 -63 V
1509 251 M
0 63 V
0 1795 R
0 -63 V
1963 251 M
0 63 V
0 1795 R
0 -63 V
2417 251 M
0 63 V
0 1795 R
0 -63 V
2872 251 M
0 63 V
0 1795 R
0 -63 V
3326 251 M
0 63 V
0 1795 R
0 -63 V
600 251 M
2817 0 V
0 1858 V
-2817 0 V
600 251 L
LT0
2114 1946 M
180 0 V
600 585 M
3 2 V
33 19 V
33 17 V
32 18 V
33 19 V
32 18 V
33 18 V
33 16 V
32 16 V
33 16 V
33 15 V
33 15 V
32 15 V
33 16 V
33 16 V
32 16 V
33 16 V
33 16 V
32 15 V
33 14 V
33 14 V
32 14 V
33 13 V
33 14 V
32 13 V
33 14 V
33 15 V
33 15 V
32 14 V
33 15 V
32 14 V
33 15 V
33 14 V
32 14 V
33 13 V
33 13 V
33 13 V
32 13 V
33 12 V
33 13 V
32 13 V
33 13 V
33 13 V
32 14 V
33 13 V
32 14 V
33 14 V
33 14 V
33 14 V
32 14 V
33 14 V
33 15 V
32 15 V
33 15 V
33 15 V
32 15 V
33 15 V
33 15 V
32 15 V
33 15 V
33 14 V
32 15 V
33 12 V
33 12 V
33 12 V
32 12 V
33 11 V
33 12 V
32 11 V
33 12 V
32 11 V
33 12 V
33 12 V
33 12 V
32 13 V
33 13 V
33 13 V
32 14 V
33 14 V
33 15 V
32 16 V
33 16 V
33 17 V
32 17 V
33 19 V
33 19 V
32 21 V
LT1
2114 1846 M
180 0 V
600 579 M
3 2 V
33 20 V
33 21 V
32 19 V
33 19 V
32 18 V
33 18 V
33 17 V
32 15 V
33 15 V
33 13 V
33 14 V
32 13 V
33 15 V
33 13 V
32 14 V
33 13 V
33 13 V
32 12 V
33 10 V
33 10 V
32 9 V
33 9 V
33 9 V
32 8 V
33 9 V
33 9 V
33 9 V
32 9 V
33 8 V
32 8 V
33 8 V
33 8 V
32 6 V
33 7 V
33 6 V
33 5 V
32 6 V
33 5 V
33 4 V
32 5 V
33 4 V
33 3 V
32 4 V
33 3 V
32 3 V
33 2 V
33 3 V
33 3 V
32 2 V
33 2 V
33 3 V
32 3 V
33 4 V
33 3 V
32 3 V
33 2 V
33 3 V
32 2 V
33 2 V
33 1 V
32 2 V
33 -1 V
33 -2 V
33 -2 V
32 -2 V
33 -2 V
33 -3 V
32 -3 V
33 -2 V
32 -3 V
33 -2 V
33 -3 V
33 -2 V
32 -2 V
33 -1 V
33 -1 V
32 0 V
33 0 V
33 0 V
32 2 V
33 2 V
33 2 V
32 4 V
33 5 V
33 5 V
32 7 V
LT2
2114 1746 M
180 0 V
600 597 M
3 1 V
33 10 V
33 9 V
32 14 V
33 17 V
32 17 V
33 18 V
33 16 V
32 14 V
33 14 V
33 13 V
33 13 V
32 13 V
33 13 V
33 14 V
32 14 V
33 13 V
33 13 V
32 12 V
33 11 V
33 10 V
32 10 V
33 10 V
33 10 V
32 9 V
33 10 V
33 11 V
33 10 V
32 11 V
33 10 V
32 10 V
33 10 V
33 10 V
32 9 V
33 8 V
33 8 V
33 8 V
32 8 V
33 7 V
33 8 V
32 7 V
33 7 V
33 8 V
32 8 V
33 7 V
32 8 V
33 8 V
33 7 V
33 8 V
32 7 V
33 8 V
33 8 V
32 8 V
33 8 V
33 8 V
32 7 V
33 8 V
33 7 V
32 7 V
33 6 V
33 6 V
32 6 V
33 4 V
33 2 V
33 2 V
32 2 V
33 2 V
33 2 V
32 1 V
33 1 V
32 2 V
33 1 V
33 2 V
33 2 V
32 2 V
33 3 V
33 3 V
32 3 V
33 4 V
33 5 V
32 6 V
33 6 V
33 7 V
32 8 V
33 9 V
33 10 V
32 11 V
stroke
grestore
end
showpage
}
\put(2054,1746){\makebox(0,0)[r]{${\cal S}$ with $V_{14}$}}
\put(2054,1846){\makebox(0,0)[r]{${\cal S}$ with Reid68}}
\put(2054,1946){\makebox(0,0)[r]{${\cal S}$ with Paris}}
\put(2008,21){\makebox(0,0){$n$ (fm$^{-3}$)}}
\put(100,1180){%
\special{ps: gsave currentpoint currentpoint translate
270 rotate neg exch neg exch translate}%
\makebox(0,0)[b]{\shortstack{${\cal S}(n)$ (MeV)}}%
\special{ps: currentpoint grestore moveto}%
}
\put(3326,151){\makebox(0,0){0.7}}
\put(2872,151){\makebox(0,0){0.6}}
\put(2417,151){\makebox(0,0){0.5}}
\put(1963,151){\makebox(0,0){0.4}}
\put(1509,151){\makebox(0,0){0.3}}
\put(1054,151){\makebox(0,0){0.2}}
\put(600,151){\makebox(0,0){0.1}}
\put(540,2109){\makebox(0,0)[r]{80}}
\put(540,1844){\makebox(0,0)[r]{70}}
\put(540,1578){\makebox(0,0)[r]{60}}
\put(540,1313){\makebox(0,0)[r]{50}}
\put(540,1047){\makebox(0,0)[r]{40}}
\put(540,782){\makebox(0,0)[r]{30}}
\put(540,516){\makebox(0,0)[r]{20}}
\put(540,251){\makebox(0,0)[r]{10}}
\end{picture}
     \caption{Symmetry energy ${\cal S}$ as function of density $n$
      for the Paris, Reid and Argonne $V_{14}$ potentials.}
     \label{fig:sec1fig1}
\end{figure}

This difference could be caused  either by the method used to solve the
many-body problem or by the NN interaction employed. 
Therefore, as a first step, we
have recalculated the symmetry energy using the BHF approach described below
and applied various realistic NN interactions. Results for these symmetry
energies ${\cal S}(n)$ as a function of density $n$ are displayed in 
Fig.\ \ref{fig:sec1fig1} for models of the NN interaction such as the 
Reid \cite{reid68}, Argonne $V_{14}$ \cite{wsa84} and Paris 
potentials \cite{paris80}. Using the Argonne $V_{14}$ potential, the BHF
calculation yields a symmetry energy which is almost constant for densities 
above twice nuclear matter density, consistent with results from
corresponding variational calculations. The Reid potential leads to a
symmetry energy which, at high densities, is even lower than the Argonne
result, while 
the symmetry energy 
from the Paris potential 
increases almost linearly with density. 
These results indicate that the differences
in the prediction of the symmetry energy at high densities are 
not caused by the
many-body method used, but rather originates from the various models
used for the NN interaction.

A symmetry energy derived from the Reid or Argonne potential
has in turn the consequence that the 
direct URCA process can never occur, or if
it occurs, it starts at very high densities. For the $V_{14}$ 
in a BHF calculation, it starts at a density of $1.5$ fm$^{-3}$.
This should be contrasted to the results 
with the Paris potential where
the direct URCA process occurs at a density of $0.9$ fm$^{-3}$.

However, when 
comparing these results, one should note that these potentials are
not phase-shift equivalent, i.e., they do not predict
exactly the same NN phases.
Furthermore, the predicted phase shifts do not agree accurately
with modern phase shift analyses, and the fit of the NN data
is typically 
$\chi^2$/datum $\approx 3$. During
the last years, progress has been 
made not only in the accuracy and the consistency
of the phase-shift analysis, but also in the fit of realistic NN potentials to
these data. As a result, several new NN potentials have been
constructed
which fit the world data for pp and np
scattering below 350 MeV with high precision. 
Potentials like the recent CD-Bonn
\cite{cdbonn}, the Argonne $V_{18}$ \cite{v18}
or the new Nijmegen \cite{nim} potentials yield a $\chi^2$/datum
of about 1 and may be
called phase-shift equivalent.

The aim of this paper is therefore 
to investigate whether these modern
phase equivalent potentials also predict differences in the symmetry energy
similar to those shown in Fig.\ \ref{fig:sec1fig1}.
Moreover, we would like to trace possible  
different trends in the symmetry energy back to features of the 
NN potentials and their contributions in various partial
waves. 
In order to do so, we will try to keep 
the many-body scheme as simple as possible. Here
we will employ a non-relativistic BHF approach to evaluate
the symmetry energy. In order to prepare the ground for studies
of infinite matter, we present in the next section 
some results for the scattering
matrix, while 
in section \ref{sec:sec3} we discuss the symmetry energy
for infinite matter.
Concluding remarks are presented in section \ref{sec:sec4}.

\section{Phase-shift equivalent NN potentials}
\label{sec:sec2}

The potentials we will employ here are the recent models
of the Nijmegen group \cite{nim}, the Argonne $V_{18}$ potential \cite{v18}
and the charge-dependent Bonn potential (CD-Bonn \cite{cdbonn}).
In 1993, the Nijmegen group
presented a phase-shift analysis of all proton-proton and neutron-proton
data below $350$ MeV with a $\chi^2$ per datum of $0.99$
for 4301 data entries. The above potentials have all been constructed
based on these data. The CD-Bonn potential has 
a $\chi^2$ per datum of $1.03$ and the same is true for the Nijm-I, Nijm-II
and Reid93 potential
versions of the Nijmegen group \cite{nim}. The new Argonne
potential $V_{18}$ \cite{v18} has a $\chi^2$ per datum of $1.09$. 

Although all these potentials predict almost identical phase shifts,
their mathematical structure is quite different. The Argonne 
potential, the Nijm-II and the Reid93  potentials 
are non-relativistic potential models defined in terms of local potential
functions, which are attached to various (non-relativistic)
operators of the spin, isospin
and/or angular momentum operators of the interacting pair of nucleons.
Such approaches to the NN potential have traditionally been quite
popular since they
are numerically easy to use in configuration space calculations.
The Nijm-I model is similar to the Nijm-II model, but it  
includes also a $\bf p^2$ term,
see Eq.\ (13) of Ref.\ \cite{nim},  
which may be interpreted as a non-local contribution to the central
force.
The CD-Bonn
potential is based on the relativistic meson-exchange model
of Ref.\ \cite{mac89} which is non-local and cannot be
described correctly in terms of local potential functions.
However, it is represented most conveniently in terms of partial waves.

\begin{figure}[htbp]
\begin{center}
{\centering
\mbox{\psfig{figure=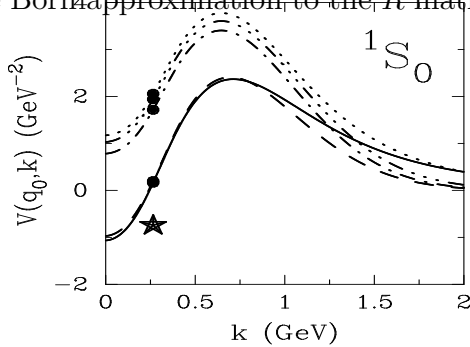,height=6cm,width=10cm}}
}
\caption{Matrix elements $V(q_0,k)$ for the $^1S_0$ partial wave for the 
CD-Bonn (solid line),
Nijm-I (dashed), Nijm-II (dash-dot), Argonne $V_{18}$
(dash-triple-dot) and Reid93 (dotted) potentials.
The diagonal matrix elements with $k=q_0=265$ MeV/c (equivalent
to $T_{lab}=150$ MeV) are marked by a solid dot.
The corresponding matrix element of the full scattering $R$-matrix is marked 
by the star.}
\label{fig:sec2fig1}
\end{center}
\end{figure}

\begin{figure}[htbp]
\begin{center}
{\centering
\mbox{\psfig{figure=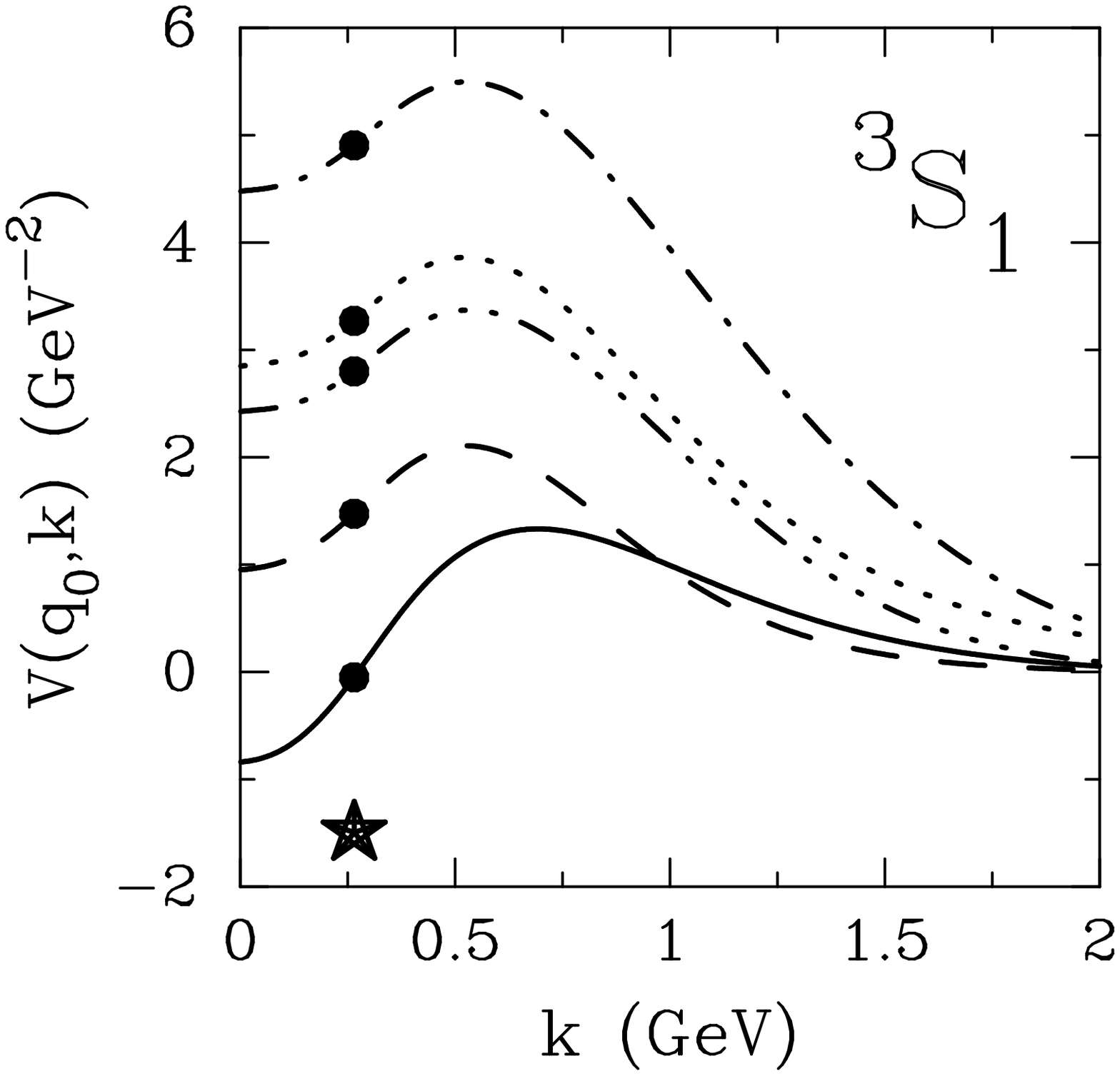,height=6cm,width=10cm}}
}
{\centering
\mbox{\psfig{figure=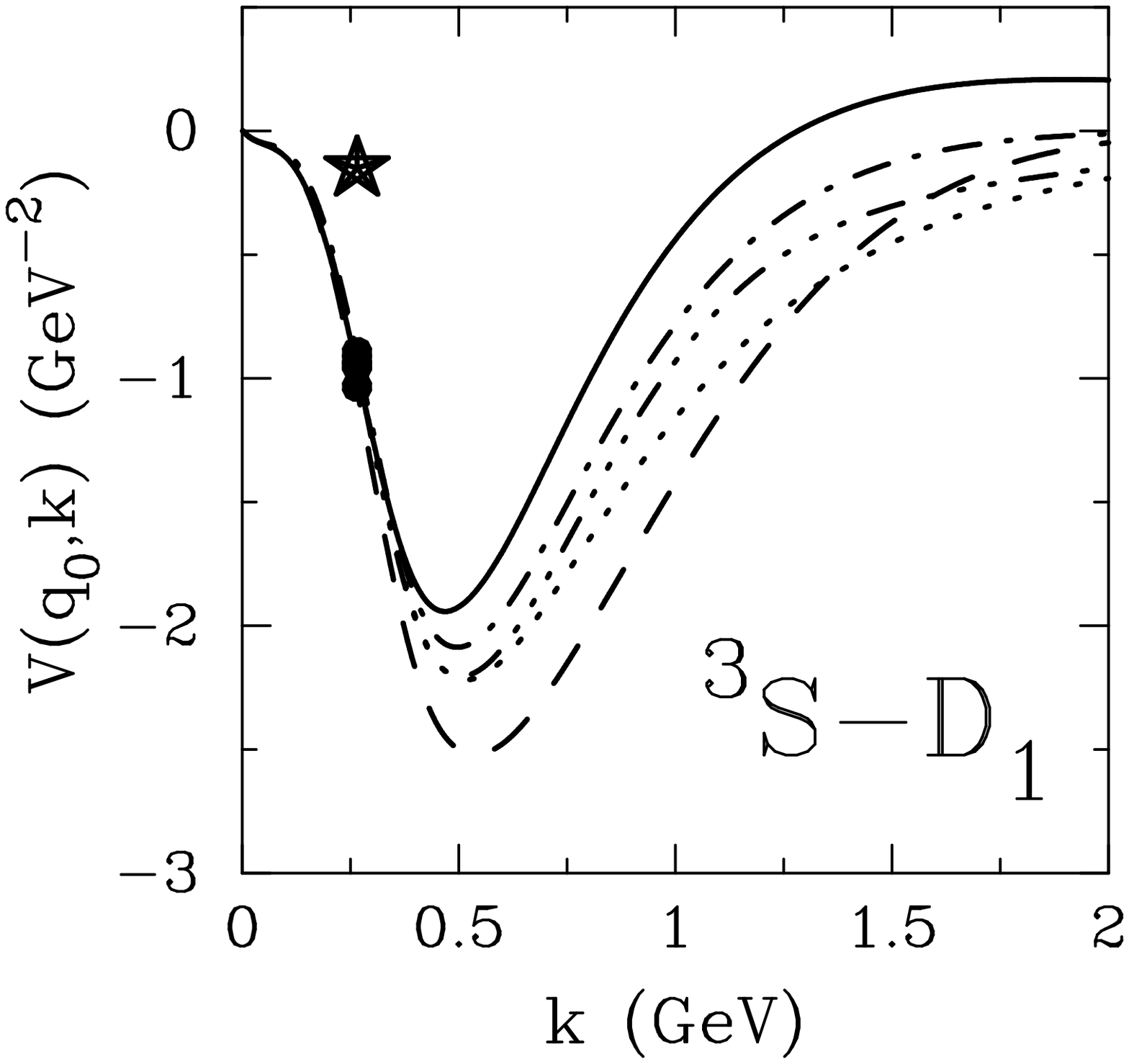,height=6cm,width=10cm}}
}
\caption{Matrix elements for the $^3S_1$ and $^3S_1$--$^3D_1$ 
potentials.
Legend as in Fig.\ 2.}
\label{fig:sec2fig2}
\end{center}
\end{figure}

\begin{figure}[htbp]
\begin{center}
{\centering
\mbox{\psfig{figure=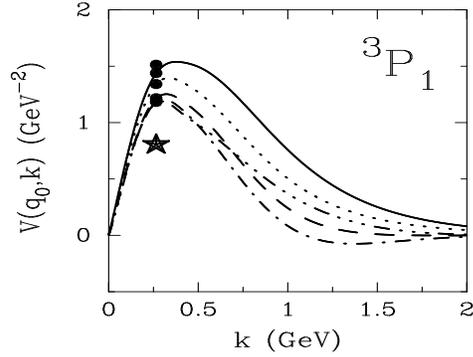,height=6cm,width=10cm}}
}
\caption{Matrix elements for the $^3P_1$ partial wave.
Legend as in Fig.\ 2.}
\label{fig:sec2fig3}
\end{center}
\end{figure}

For a given NN potential $V$, the $R$-matrix (or $K$-matrix) for free-space
two-nucleon scattering is obtained from the Lippmann-Schwinger
equation, which reads in the center-of-mass (c.m.) system 
and in partial-wave decomposition
\begin{equation}
     R^{JST}_{l'l}({ q'}, { q};E) = V^{JST}_{l'l}({ q'}, { q}) + 
     \sum_{l''} {\cal P}\int_0^\infty k^2 dk V^{JST}_{l'l''}({ q'}, { k})
     \frac{M}{ME-k^2}
      R^{JST}_{l''l}({ k}, { q};E).
\label{eq:tmatrix}
\end{equation}
The label $l$ represents the orbital angular momentum of the relative
motion, $J$ is the total angular momentum, $S$ the spin and $T$ the isospin.
Relative momenta are given by $q,q',k$, while the energy of the interacting
particles is denoted by $E$. Notice that the principle value integral (${\cal
P}$) for the integration over the intermediate momenta $k$  extends from
zero to infinity. 
The phase-shifts for a given partial wave can be calculated from the on-shell
matrix element of $R$, which is obtained by setting
\begin{equation}
q = q' = q_0 \qquad \mbox{and} \qquad E=\frac{q_0^2}{M} \label{eq:onshel}
\end{equation}
Since all NN interactions considered in this paper reproduce the same
phase-shifts, the corresponding on-shell matrix elements of $R$
calculated from these various potentials 
are identical as well. As an example, we
display the value of these $R$ matrix elements in Figs.\ \ref{fig:sec2fig1} - 
\ref{fig:sec2fig3} for NN scattering with $T_{lab}=2E=150$ MeV and a
corresponding value of $q_0=265$ MeV/c. This $R$ matrix element 
is denoted by a star and is identical 
for all 5 potentials under consideration. 
What is different, is the contribution of 
the first term on the right-hand side 
of Eq.(\ref{eq:tmatrix}) to this value of $R$, which is the Born approximation 
to $R$, and the contributions of second and higher order in $V$, which are
contained in the second term on the right-hand side of Eq.(\ref{eq:tmatrix}).
This is demonstrated in Figs.\ \ref{fig:sec2fig1} - \ref{fig:sec2fig3}, which
display the matrix elements 
$V^{JST}_{l'l}({ q_0}, { k})$ 
for the different bare
potentials as a function of $k$. The diagonal matrix element
$V^{JST}_{l'l}({ q_0}, { q_0})$, 
which represents the Born approximation to the $R$ matrix,
is denoted by a solid dot.

Looking e.g., at the results for the $^3S_1$ channel displayed in Fig.\
\ref{fig:sec2fig2} we see that the Nijm-II potential is a rather ``hard''
potential in the sense that the diagonal matrix element of the bare potential
is very repulsive. In order to obtain the attractive matrix element of $R$,
which yields the empirical phase shift, a lot of attraction has to be supplied
from the contributions to $R$ which are beyond the Born-approximation. On the
other hand, the CD-Bonn potential is rather ``soft'' in this channel, 
implying that much less
attraction is needed from the terms of higher order in $V$ in Eq.\
(\ref{eq:tmatrix}), to obtain the empirical value of $R$. 

Similar features can also be observed in the $^1S_0$ and $^3P_1$ partial waves
in Figs.\ 
\ref{fig:sec2fig1} and \ref{fig:sec2fig3}, respectively. Different potentials
turn out to be the ``softest'' potential in the various partial waves.
It should be noted
that the differences between  the various potentials are getting smaller with
increasing $l$.  For $D$-waves and higher partial waves
the various potentials are almost equal. There are still differences
for $P$-waves\footnote{The original Nijm-II potential yielded
an unphysical bound state for the $^1P_1$ wave, as pointed out by
Nogga {\em et al.} \cite{nhkg97}. In the present work, the updated
Nijm-II potential is used \cite{stoks97}.}.

So what are the differences between the various NN potentials discussed so far,
which might be responsible for the differences displayed in Figs.\  
\ref{fig:sec2fig1} - \ref{fig:sec2fig3}? The NN potentials denoted by Reid93
and Nijm-II \cite{nim} are purely local potentials in the sense that 
they consider the
local form of the One-Pion-Exchange potential for the long-range part and
parametrize the contributions of medium and short-range in terms of local
functions, depending on the distance between the interacting nucleons,
multiplying a set of spin-isospin operators. The same is also true for the
Argonne $V_{18}$ potential \cite{v18}. The NN potential denoted as Nijm-I
\cite{nim} also contains the local representation of the One-Pion-Exchange part
but keeps track of non-localities of the medium- and short-range 
central-force components,
which one obtains by evaluating these contributions from a One-Boson-Exchange
model in terms of second-order Feynman diagrams. The CD-Bonn potential on the
other hand \cite{cdbonn} has been evaluated completely within this scheme. It
has been shown \cite{cdbonn} that  ignoring the non-localities in the 
One-Pion-Exchange part leads 
to a larger tensor component in the bare potential.
This may also be the
origin of the fact that the CD-Bonn potential yields a smaller D-state
probability for the Deuteron ($P_D = 4.83 \%$) than the other potentials 
($P_D \approx 5.6 \%$). 

Returning to Figs.\ \ref{fig:sec2fig1} - \ref{fig:sec2fig3} we see
that for the $^1S_0$ partial wave, which is not affected by the tensor
component, the NN interactions CD-Bonn and Nijm-I, which both account for
non-localities, exhibit rather similar results for the bare potential, while
the pure local potentials do require more attraction from the non-Born terms to
get to the empirical value of $R$. 
In the case of the $^3S_1$ channel we observe
a difference also between the CD-Bonn and the Nijm-I potential, which should be
related to the fact that the CD-Bonn potential includes non-local effects also 
in the One-Pion-Exchange contribution. As it has been demonstrated in
\cite{cdbonn} the neglect of non-local components in the One-Pion-Exchange term
increases the non-diagonal matrix elements of $V$ in the coupling channel
$^3S_1$--$^3D_1$ (cf.\ Fig.~3). 
This implies that the contributions to $R$ in the $^3S_1$
partial wave, which are of second and higher order in $V$ (see Eq.
\ref{eq:tmatrix}), give rise to more attraction, when local approximations to
One-Pion-Exchange term are considered. Consequently, the diagonal matrix
elements of $V$ in this channel must be less attractive to obtain the same
matrix element for $R$. In the language used above this means that the
potential which accounts for the non-locality of the One-Pion-Exchange (the
CD-Bonn) is ``softer'' than the other potentials (see Fig.\
\ref{fig:sec2fig2}). 

\section{Symmetry energy in infinite matter}\label{sec:sec3}

The many-body method we will employ in deriving the symmetry 
energy and the relevant proton fractions is a rather simple one, the 
non-relativistic Brueckner-Hartree-Fock (BHF) method with a continuous
single-particle spectrum \cite{mahaux85}. Here it will
serve us as a mere tool to investigate various NN potentials so that
the discrepancies observed can be retraced to the  NN potentials only.
This is also the main aim of this work, as discussed
in the introduction as well. Moreover, the $G$-matrix offers 
a rather direct link to the $R$-matrix discussed in the previous section.
Therefore, as discussed below, eventual differences
between various potentials in a finite medium should be easily retraced
to e.g.,
Figs.\ \ref{fig:sec2fig1}-\ref{fig:sec2fig3} and the structure 
of the $R$-matrix.
 
Before we proceed, we feel however that 
a disclaimer is necessary. We do not, by obvious reasons, believe
that a non-relativistic BHF scheme offers the most realistic approach 
to matter at
densities above nuclear matter saturation density $n=0.17$ fm$^{-3}$.
Other many-body contributions such as three-hole line contributions
could be important in order to reproduce the empirical data. Although
recent investigations by Baldo et al.\ \cite{trebal} indicate that the 
BHF approach with a continuous choice for the single-particle 
spectrum accounts
for three-hole line contributions in nuclear matter (see also the work of
Day \cite{day81}), further many-body terms as included in the 
recent variational
calculation of Akmal and Pandharipande \cite{ap97}. 
Moreover, accounting only for nucleonic degrees of freedom 
should be viewed as nothing but a first approximation. There is no
reason why other baryons should not be present at higher densities.

The energy per particle ${\cal E}(n,\chi_p)$ within the BHF scheme is 
given in terms of the 
so-called reaction matrix $G$. The latter is obtained by solving the
Bethe-Goldstone equation for various proton fractions $\chi_p$ \cite{ehobo96}
\begin{equation}
       G(\omega ,\chi_p )=V+V\frac{Q(\chi_p)}{\omega - H_0}G(\omega ,\chi_p),
       \label{eq:bg}
\end{equation}
where $\omega$ is the unperturbed energy of the interacting  nucleons,
$V$ is the free NN potential, $H_0$ is the unperturbed energy of the
intermediate scattering states,
and $Q(\chi_p)$ is the Pauli
operator preventing scattering into occupied states.
Only ladder diagrams with 
intermediate two-particle states are included in Eq.\ 
(\ref{eq:bg}). The structure of the Bethe-Goldstone equation can 
then be directly
compared to the $R$ matrix for free NN scattering, Eq.~(\ref{eq:tmatrix}). Note
that in (\ref{eq:bg}) we have suppressed the quantum numbers for the various
partial waves as well as the integral over the intermediate two-particle states
in order to simplify the notation. 

It is obvious that one expects the matrix elements of $G$ to be rather close to
those of $R$ with only small deviations. These deviations originate from two
effects which reduce the contributions of second and higher order in $V$ to the
$G$-matrix as compared to their contributions to $R$. One is the so-called
Pauli quenching effect: 
the Pauli operator $Q$ in (\ref{eq:bg}) restricts the
intermediate particle states to states above the Fermi energy. 
The second one is
the dispersive effect: the energy denominators in (\ref{eq:bg}) are defined in
terms of the single-particle energies of nucleons in the medium while the
corresponding denominators of (\ref{eq:tmatrix}) are differences between the
energies of free nucleons. Since the absolute values for the energy differences
between nucleons, which feel the mean field of the nuclear system, are larger
than the energy differences between the kinetic energies, also this dispersive
correction reduces the attractive contributions of the non-Born terms. 

As a result, the matrix elements of $G$ tend to be less attractive
than the corresponding matrix elements of $R$. 
This difference is due to the two
quenching mechanisms which we just discussed. Since it originates from the
quenching of the non-Born terms, it is smaller for a ``soft'' potential (using
the terminology of the previous section) 
since the contribution of the non-Born
term is weaker in this case.  
With the $G$-matrix we can calculate the total energy per nucleon 
\begin{equation}
        {\cal E}(n,\chi_p)={\cal T}+{\cal U}(n,\chi_p),
        \label{eq:g_bhf}
\end{equation}
with the kinetic energy
\begin{equation}
{\cal T} = \frac{3 k_{Fp}^5 + 3 k_{Fn}^5}{10 m\, k_F^3}\label{eq:tkin}
\end{equation}
where the total Fermi momentum $k_F$ and the Fermi momenta $k_{Fp}$,
$k_{Fn}$ for protons and neutrons are related to the total nucleon density
$n$ by
\begin{eqnarray}
n & = & \frac{2}{3\pi^2} k_F^3 \nonumber \\
& = & \chi_p n + (1-\chi_p) n \nonumber \\
& = & \frac{1}{3\pi^2} k_{Fp}^3 + \frac{1}{3\pi^2} k_{Fn}^3 \label{eq:densi}
\end{eqnarray}
$\chi_p$ denotes the proton fraction and corresponds to the ratio of protons as
compared to the total nucleon number ($Z/A$). The contribution of the potential
energy ${\cal U}$ to the total energy per particle can be written
\begin{equation}
        {\cal U}(n,\chi_p)=\frac{1}{2n}\frac{1}{(2\pi)^6}\sum _{a,b=(pn)} 
        \int_0^{k_{Fa}} d^3k_a \int_0^{k_{Fb}} d^3k_b
         \langle k_ak_b\vert G(\omega =
         \epsilon _a(k_a) + \epsilon _b(k_b))\vert k_ak_b\rangle,
        \label{eq:upot}
\end{equation}
The  single-particle energies for protons and neutrons are denoted
by $\epsilon_a$. For a given density $n$ and proton fraction $\chi_p$ they 
depend on the momentum $k$ of the nucleon  and are determined self-consistently
using the BHF theory. The integrals in (\ref{eq:upot}) can be decomposed into
contributions of various partial waves. In the limit of pure neutron matter
only those partial waves contribute, in which the pair of interacting nucleons
is coupled to isospin $T=1$. Due to the antisymmetry of the matrix elements
this implies that only partial waves with even values for the sum $l+S$, like
$^1S_0$, $^3P_0$ etc.\ need to be considered in this case. For proton fractions
different from zero, in particular the case of symmetric nuclear matter, also
the other partial waves, like $^3S_1 - ^3D_1$ and $^1P_1$ contribute.
See e.g.\ Ref.\ \cite{hko95} for further details.

\begin{figure}[htbp]
     \input{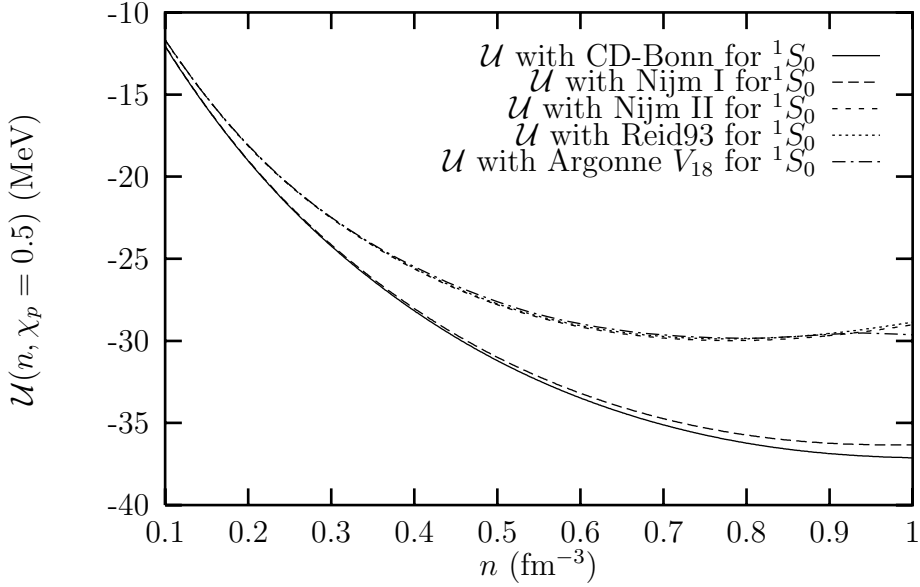}
     \caption{Potential energy per particle ${\cal U}$ 
              for symmetric nuclear matter originating
              from the $^1S_0$ partial wave only as a function of density
              $n$. Results are shown for the CD-Bonn potential 
              \protect\cite{cdbonn}, the new Argonne $V_{18}$ potential
              \protect\cite{v18} and the three  recent versions of the Nijmegen
              group \protect\cite{nim}.}
     \label{fig:sec3fig1}
\end{figure}

Since the contribution of the kinetic energy ${\cal T}$ to the total energy is
independent on the NN interaction chosen, we will restrict the following
discussion to the potential energy per nucleon ${\cal U}$ and explore the
contribution of the various partial waves to it. As a first example we display
in Fig.\ \ref{fig:sec3fig1} the contribution of the $^1S_0$ partial wave to the
potential energy per nucleon ${\cal U}$ of symmetric nuclear matter as a 
function of density $n$. The  potentials employed are the CD-Bonn potential 
\cite{cdbonn}(solid line), the three Nijmegen potentials, Nijm-I (long 
dashes), Nijm-II (short dashes) and Reid93 (dotted line)\cite{nim} and 
the Argonne $V_{18}$ \cite{v18} (dot-dashed line). 

One observes that the results for the CD-Bonn potential are essentially 
identical to those of the Nijm-I. Both potentials yield substantially more 
attraction in this particular partial wave than the other three 
considered here. 
This is in complete agreement with the observations made in the previous
section (see Fig.\ \ref{fig:sec2fig1}). There we observed that the CD-Bonn 
and the Nijm-I potential, the two potentials accounting for non-local effects 
in the medium- and short-range components of the central force 
are ``softer'' in this partial
wave than the other two. 
This means that the quenching effects, which reduce the
attraction of the non-Born terms in $G$ as compared to $R$ (as discussed above)
are less efficient for these two potentials as compared to Nijm-II, Reid93
and Argonne $V_{18}$.
Consequently the CD-Bonn potential and the Nijm-I yield more attraction.
At a density of e.g., 
$0.6$ fm$^{-3}$, we get
a contribution to the potential energy from the 
$^1S_0$ channel
of $-33.4$, $-33.2$ , $-29.2$, $-29.1$ and $-29.1$ MeV
for the CD-Bonn, Nijm I, Nijm II, Reid93 and Argonne potentials, 
respectively.

\begin{figure}[htbp]
     \input{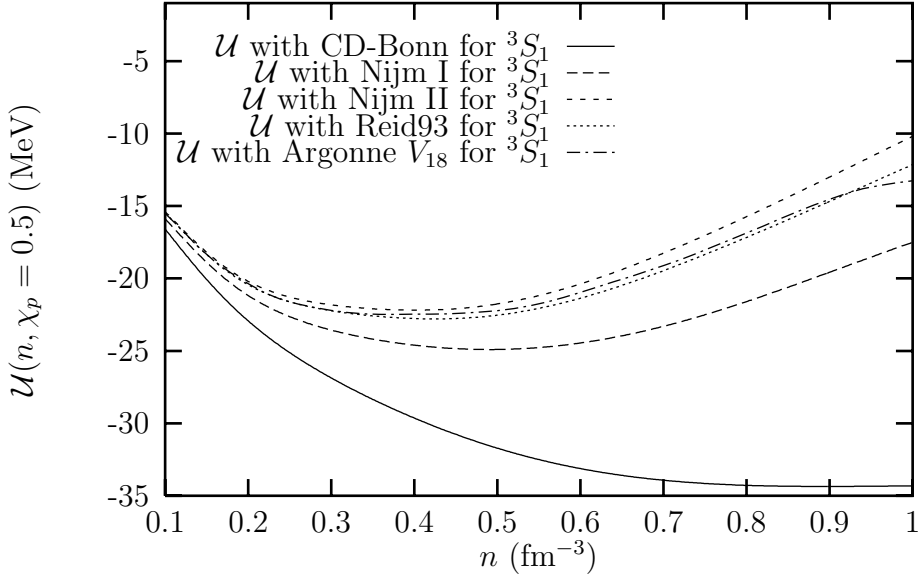}
     \caption{Potential energy per particle ${\cal U}$ 
              for symmetric nuclear matter originating from
              the $^3S_1$ partial wave only.}
     \label{fig:sec3fig2}
\end{figure}

Similar observations can also be made for the contribution of the $^3S_1$
partial wave to the binding energy per nucleon displayed in Fig.\ 
\ref{fig:sec3fig2}. The ``softer'' the potential in this channel (compare 
Fig.\ \ref{fig:sec2fig2}), the more attraction 
is obtained. 
Although the differences we observed in the previous 
figure for the $^1S_0$ wave are present here as well, 
this partial wave is
particularly sensitive to the tensor component of the NN interaction, which
originates mainly from the One-Pion-Exchange contribution. Due to these strong
tensor correlations, the non-Born terms in Eqs.\ (\ref{eq:tmatrix}) and
(\ref{eq:bg}) are more important in the $^3S_1$ than in the $^1S_0$ channel.
Therefore, 
the differences in the binding energy contribution predicted from the
different interactions are particularly large in this channel.
A potential model  with a weak tensor force
yields more attraction in a nuclear system than a 
potential with a strong tensor force. This is clearly seen in
Fig.\ \ref{fig:sec3fig2} where we study the contribution
from the $^3S_1$ partial wave in the 
contribution to the potential energy ${\cal U}$.
At a density of 
$0.6$ fm$^{-3}$, we get
a contribution to the potential energy from the 
$^3S_1$ channel
of $-33$, $-24$ , $-20.4$, $-21.3$ and $-20.1$ MeV
for the CD-Bonn, Nijm I, Nijm II, Reid93 and Argonne potentials, 
respectively, a difference of more
than 50\% between the CD-Bonn potential 
and the Nijm II, Reid93 and Argonne potentials. 
However, it is not only the tensor which plays a significant
role in explaining the differences seen in Figs.\ \ref{fig:sec3fig1}
and \ref{fig:sec3fig2}. 
The Nijm I potential has a non-local term to the central
force, a term which explains the differences of 
approximately 20\% between the Nijm I potential 
and the Nijm II and Reid93 potentials. This is the same mechanism
discussed in connection with Fig.\ \ref{fig:sec3fig1}.
The non-localities included in the tensor force of the CD-Bonn
potential are in turn responsible for the further difference of 9 MeV
between the CD-Bonn potential and the Nijm I potential. 
The differences discussed in Fig.\ \ref{fig:sec2fig2} and the 
way they affect the integral term in the $G$-matrix (and $T$)
are clearly seen in Fig.\ \ref{fig:sec3fig2}.

\begin{figure}[htbp]
     \input{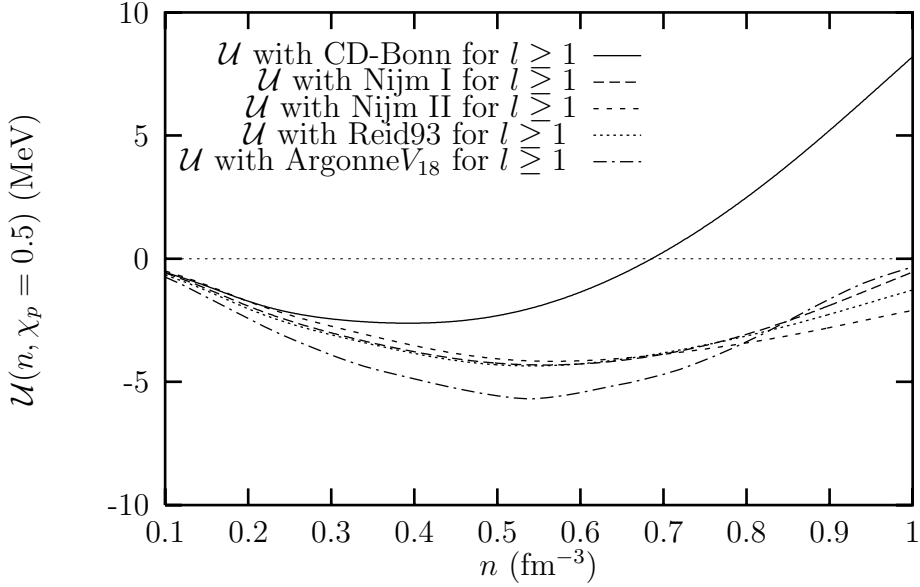}
     \caption{Potential energy per particle ${\cal U}$ 
              for symmetric nuclear matter originating from
              partial waves with $l\ge 1$.}
     \label{fig:sec3fig3}
\end{figure}

Thus, Figs.\ \ref{fig:sec3fig1} and \ref{fig:sec3fig2}
demonstrate in a clear way the importance of the tensor force
and non-local terms in the construction of the nucleon-nucleon
force. 
In addition, non-negligible differences seen in $P$-waves
are reflected in different behaviors for partial waves
with $l \geq 1$. 
In Fig.\ \ref{fig:sec3fig3} we display the contributions from partial
waves with
$l\geq 1$. In the $P$ channels, 
the CD-Bonn potential predicts less attraction
than the other potentials. 
This in line with the observation made in the 
previous section (see 
Fig.\ \ref{fig:sec2fig3}) that for these partial waves, the Nijmegen potentials
are ``softer''. 
Note also that the Argonne $V_{18}$ slightly deviates from the
Nijmegen potentials. The difference can mainly be retraced to different
contributions from $P$-waves. This difference, together with that
observed in the $^1S_0$ channel of Fig.\ \ref{fig:sec3fig1}, will be important
in explaining the behavior of the new Argonne potential in neutron matter.
For partial waves with $l\geq 2$, the results
differ by some few keV only.

\begin{figure}[htbp]
\setlength{\unitlength}{0.1bp}
\special{!
/gnudict 40 dict def
gnudict begin
/Color false def
/Solid false def
/gnulinewidth 5.000 def
/vshift -33 def
/dl {10 mul} def
/hpt 31.5 def
/vpt 31.5 def
/M {moveto} bind def
/L {lineto} bind def
/R {rmoveto} bind def
/V {rlineto} bind def
/vpt2 vpt 2 mul def
/hpt2 hpt 2 mul def
/Lshow { currentpoint stroke M
  0 vshift R show } def
/Rshow { currentpoint stroke M
  dup stringwidth pop neg vshift R show } def
/Cshow { currentpoint stroke M
  dup stringwidth pop -2 div vshift R show } def
/DL { Color {setrgbcolor Solid {pop []} if 0 setdash }
 {pop pop pop Solid {pop []} if 0 setdash} ifelse } def
/BL { stroke gnulinewidth 2 mul setlinewidth } def
/AL { stroke gnulinewidth 2 div setlinewidth } def
/PL { stroke gnulinewidth setlinewidth } def
/LTb { BL [] 0 0 0 DL } def
/LTa { AL [1 dl 2 dl] 0 setdash 0 0 0 setrgbcolor } def
/LT0 { PL [] 0 1 0 DL } def
/LT1 { PL [4 dl 2 dl] 0 0 1 DL } def
/LT2 { PL [2 dl 3 dl] 1 0 0 DL } def
/LT3 { PL [1 dl 1.5 dl] 1 0 1 DL } def
/LT4 { PL [5 dl 2 dl 1 dl 2 dl] 0 1 1 DL } def
/LT5 { PL [4 dl 3 dl 1 dl 3 dl] 1 1 0 DL } def
/LT6 { PL [2 dl 2 dl 2 dl 4 dl] 0 0 0 DL } def
/LT7 { PL [2 dl 2 dl 2 dl 2 dl 2 dl 4 dl] 1 0.3 0 DL } def
/LT8 { PL [2 dl 2 dl 2 dl 2 dl 2 dl 2 dl 2 dl 4 dl] 0.5 0.5 0.5 DL } def
/P { stroke [] 0 setdash
  currentlinewidth 2 div sub M
  0 currentlinewidth V stroke } def
/D { stroke [] 0 setdash 2 copy vpt add M
  hpt neg vpt neg V hpt vpt neg V
  hpt vpt V hpt neg vpt V closepath stroke
  P } def
/A { stroke [] 0 setdash vpt sub M 0 vpt2 V
  currentpoint stroke M
  hpt neg vpt neg R hpt2 0 V stroke
  } def
/B { stroke [] 0 setdash 2 copy exch hpt sub exch vpt add M
  0 vpt2 neg V hpt2 0 V 0 vpt2 V
  hpt2 neg 0 V closepath stroke
  P } def
/C { stroke [] 0 setdash exch hpt sub exch vpt add M
  hpt2 vpt2 neg V currentpoint stroke M
  hpt2 neg 0 R hpt2 vpt2 V stroke } def
/T { stroke [] 0 setdash 2 copy vpt 1.12 mul add M
  hpt neg vpt -1.62 mul V
  hpt 2 mul 0 V
  hpt neg vpt 1.62 mul V closepath stroke
  P  } def
/S { 2 copy A C} def
end
}
\begin{picture}(3600,2160)(0,0)
\special{"
gnudict begin
gsave
50 50 translate
0.100 0.100 scale
0 setgray
/Helvetica findfont 100 scalefont setfont
newpath
-500.000000 -500.000000 translate
LTa
LTb
600 251 M
63 0 V
2754 0 R
-63 0 V
600 561 M
63 0 V
2754 0 R
-63 0 V
600 870 M
63 0 V
2754 0 R
-63 0 V
600 1180 M
63 0 V
2754 0 R
-63 0 V
600 1490 M
63 0 V
2754 0 R
-63 0 V
600 1799 M
63 0 V
2754 0 R
-63 0 V
600 2109 M
63 0 V
2754 0 R
-63 0 V
600 251 M
0 63 V
0 1795 R
0 -63 V
913 251 M
0 63 V
0 1795 R
0 -63 V
1226 251 M
0 63 V
0 1795 R
0 -63 V
1539 251 M
0 63 V
0 1795 R
0 -63 V
1852 251 M
0 63 V
0 1795 R
0 -63 V
2165 251 M
0 63 V
0 1795 R
0 -63 V
2478 251 M
0 63 V
0 1795 R
0 -63 V
2791 251 M
0 63 V
0 1795 R
0 -63 V
3104 251 M
0 63 V
0 1795 R
0 -63 V
3417 251 M
0 63 V
0 1795 R
0 -63 V
600 251 M
2817 0 V
0 1858 V
-2817 0 V
600 251 L
LT0
3114 1946 M
180 0 V
631 1464 M
32 -49 V
31 -48 V
31 -48 V
32 -45 V
31 -45 V
31 -43 V
31 -42 V
32 -40 V
31 -39 V
31 -36 V
32 -35 V
31 -34 V
31 -32 V
32 -31 V
31 -29 V
31 -28 V
31 -28 V
32 -26 V
31 -25 V
31 -25 V
32 -23 V
31 -23 V
31 -22 V
32 -22 V
31 -21 V
31 -20 V
31 -20 V
32 -19 V
31 -18 V
31 -18 V
32 -17 V
31 -17 V
31 -16 V
32 -15 V
31 -15 V
31 -14 V
31 -13 V
32 -13 V
31 -11 V
31 -12 V
32 -10 V
31 -10 V
31 -10 V
32 -8 V
31 -9 V
31 -7 V
31 -7 V
32 -6 V
31 -6 V
31 -5 V
32 -4 V
31 -4 V
31 -3 V
32 -3 V
31 -3 V
31 -1 V
31 -1 V
32 -1 V
31 0 V
31 0 V
32 1 V
31 1 V
31 2 V
32 2 V
31 3 V
31 3 V
31 3 V
32 4 V
31 4 V
31 5 V
32 5 V
31 5 V
31 6 V
32 6 V
31 7 V
31 7 V
31 7 V
32 7 V
31 8 V
31 7 V
32 8 V
31 9 V
31 8 V
32 9 V
31 8 V
31 9 V
31 9 V
32 9 V
31 10 V
LT1
3114 1846 M
180 0 V
631 1489 M
32 -47 V
31 -46 V
31 -45 V
32 -43 V
31 -43 V
31 -40 V
31 -38 V
32 -37 V
31 -35 V
31 -33 V
32 -32 V
31 -29 V
31 -29 V
32 -27 V
31 -25 V
31 -25 V
31 -23 V
32 -22 V
31 -22 V
31 -20 V
32 -21 V
31 -18 V
31 -19 V
32 -18 V
31 -17 V
31 -17 V
31 -16 V
32 -15 V
31 -15 V
31 -14 V
32 -14 V
31 -13 V
31 -13 V
32 -12 V
31 -11 V
31 -10 V
31 -10 V
32 -10 V
31 -9 V
31 -8 V
32 -7 V
31 -7 V
31 -7 V
32 -6 V
31 -5 V
31 -4 V
31 -4 V
32 -4 V
31 -3 V
31 -2 V
32 -2 V
31 -1 V
31 -1 V
32 0 V
31 0 V
31 1 V
31 1 V
32 2 V
31 2 V
31 3 V
32 3 V
31 4 V
31 4 V
32 4 V
31 5 V
31 6 V
31 5 V
32 6 V
31 7 V
31 7 V
32 7 V
31 8 V
31 8 V
32 8 V
31 9 V
31 9 V
31 9 V
32 9 V
31 10 V
31 9 V
32 11 V
31 10 V
31 10 V
32 10 V
31 11 V
31 10 V
31 11 V
32 10 V
31 10 V
LT2
3114 1746 M
180 0 V
631 1521 M
32 -44 V
31 -43 V
31 -41 V
32 -41 V
31 -38 V
31 -37 V
31 -35 V
32 -33 V
31 -31 V
31 -29 V
32 -26 V
31 -25 V
31 -24 V
32 -22 V
31 -20 V
31 -20 V
31 -18 V
32 -17 V
31 -17 V
31 -15 V
32 -15 V
31 -15 V
31 -14 V
32 -13 V
31 -13 V
31 -12 V
31 -12 V
32 -11 V
31 -11 V
31 -10 V
32 -9 V
31 -9 V
31 -8 V
32 -8 V
31 -7 V
31 -6 V
31 -5 V
32 -5 V
31 -4 V
31 -3 V
32 -3 V
31 -1 V
31 -1 V
32 -1 V
31 1 V
31 0 V
31 2 V
32 2 V
31 3 V
31 3 V
32 3 V
31 5 V
31 4 V
32 6 V
31 5 V
31 7 V
31 6 V
32 7 V
31 7 V
31 7 V
32 8 V
31 8 V
31 8 V
32 9 V
31 9 V
31 9 V
31 9 V
32 10 V
31 10 V
31 10 V
32 11 V
31 11 V
31 11 V
32 11 V
31 11 V
31 12 V
31 12 V
32 12 V
31 12 V
31 13 V
32 12 V
31 13 V
31 13 V
32 13 V
31 13 V
31 13 V
31 13 V
32 13 V
31 13 V
LT3
3114 1646 M
180 0 V
631 1513 M
32 -45 V
31 -43 V
31 -43 V
32 -41 V
31 -40 V
31 -37 V
31 -36 V
32 -34 V
31 -32 V
31 -29 V
32 -28 V
31 -26 V
31 -24 V
32 -22 V
31 -21 V
31 -20 V
31 -19 V
32 -18 V
31 -17 V
31 -16 V
32 -15 V
31 -15 V
31 -15 V
32 -13 V
31 -14 V
31 -12 V
31 -13 V
32 -11 V
31 -11 V
31 -11 V
32 -10 V
31 -9 V
31 -8 V
32 -8 V
31 -7 V
31 -7 V
31 -6 V
32 -4 V
31 -4 V
31 -4 V
32 -3 V
31 -2 V
31 -1 V
32 -1 V
31 0 V
31 1 V
31 1 V
32 2 V
31 2 V
31 4 V
32 3 V
31 5 V
31 4 V
32 5 V
31 6 V
31 6 V
31 6 V
32 7 V
31 7 V
31 7 V
32 8 V
31 8 V
31 9 V
32 8 V
31 9 V
31 10 V
31 9 V
32 10 V
31 10 V
31 10 V
32 11 V
31 11 V
31 11 V
32 11 V
31 12 V
31 11 V
31 12 V
32 13 V
31 12 V
31 12 V
32 13 V
31 13 V
31 13 V
32 13 V
31 13 V
31 14 V
31 13 V
32 13 V
31 14 V
LT4
3114 1546 M
180 0 V
-2694 8 R
63 -90 V
62 -90 V
63 -82 V
62 -72 V
63 -67 V
63 -62 V
62 -54 V
63 -45 V
62 -41 V
63 -37 V
63 -34 V
62 -29 V
63 -25 V
62 -23 V
63 -20 V
63 -20 V
62 -18 V
63 -17 V
62 -14 V
63 -11 V
63 -8 V
62 -5 V
63 1 V
62 5 V
63 8 V
63 10 V
62 11 V
63 11 V
62 13 V
63 15 V
63 16 V
62 20 V
63 21 V
62 23 V
63 24 V
63 25 V
62 28 V
63 27 V
62 27 V
63 26 V
63 25 V
62 20 V
63 13 V
62 12 V
63 9 V
stroke
grestore
end
showpage
}
\put(3054,1546){\makebox(0,0)[r]{${\cal U}$ with Argonne $V_{18}$}}
\put(3054,1646){\makebox(0,0)[r]{${\cal U}$ with Reid93}}
\put(3054,1746){\makebox(0,0)[r]{${\cal U}$ with Nijm II}}
\put(3054,1846){\makebox(0,0)[r]{${\cal U}$ with Nijm I}}
\put(3054,1946){\makebox(0,0)[r]{${\cal U}$ with CD-Bonn}}
\put(2008,21){\makebox(0,0){$n$ (fm$^{-3}$)}}
\put(100,1180){%
\special{ps: gsave currentpoint currentpoint translate
270 rotate neg exch neg exch translate}%
\makebox(0,0)[b]{\shortstack{${\cal U}(n,\chi_p=0.5)$ (MeV)}}%
\special{ps: currentpoint grestore moveto}%
}
\put(3417,151){\makebox(0,0){1}}
\put(3104,151){\makebox(0,0){0.9}}
\put(2791,151){\makebox(0,0){0.8}}
\put(2478,151){\makebox(0,0){0.7}}
\put(2165,151){\makebox(0,0){0.6}}
\put(1852,151){\makebox(0,0){0.5}}
\put(1539,151){\makebox(0,0){0.4}}
\put(1226,151){\makebox(0,0){0.3}}
\put(913,151){\makebox(0,0){0.2}}
\put(600,151){\makebox(0,0){0.1}}
\put(540,2109){\makebox(0,0)[r]{-10}}
\put(540,1799){\makebox(0,0)[r]{-20}}
\put(540,1490){\makebox(0,0)[r]{-30}}
\put(540,1180){\makebox(0,0)[r]{-40}}
\put(540,870){\makebox(0,0)[r]{-50}}
\put(540,561){\makebox(0,0)[r]{-60}}
\put(540,251){\makebox(0,0)[r]{-70}}
\end{picture}
     \caption{Potential energy per particle ${\cal U}$ 
              for symmetric nuclear matter as function 
              of total baryonic density $n$.}
     \label{fig:sec3fig4}
\end{figure}
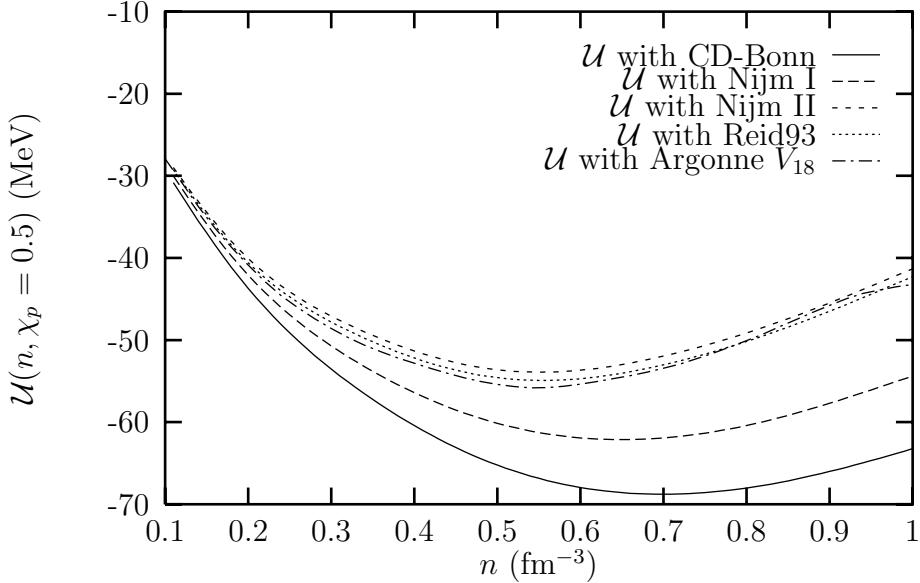

\begin{figure}[htbp]
     \input{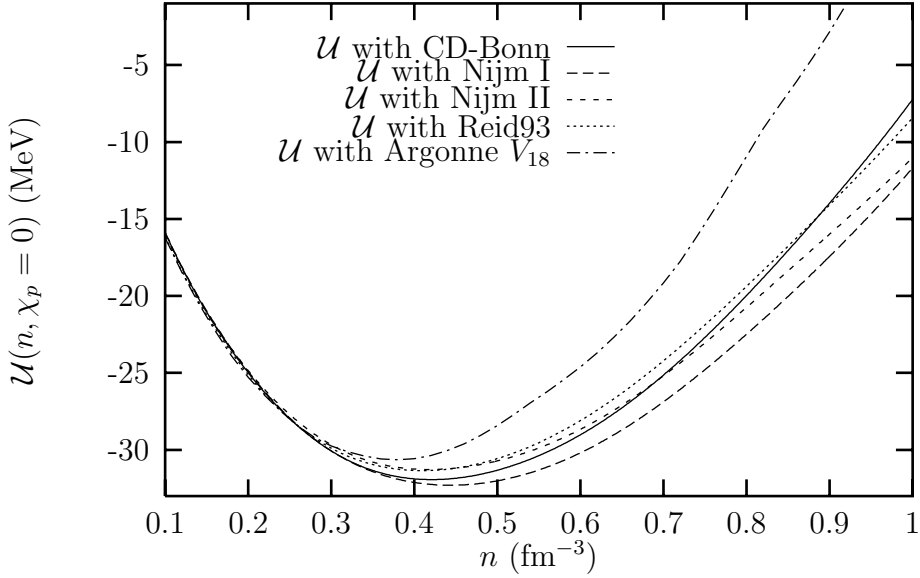}
     \caption{Potential energy per particle ${\cal U}$ 
              for pure neutron matter.}
     \label{fig:sec3fig5}
\end{figure}

Putting the contributions from 
various channels together\footnote{In our 
calculations we include all partial waves with $l < 10$.}, one obtains 
the total potential energy per nucleon ${\cal  U}$ for symmetric
nuclear matter and neutron matter. These results are displayed
in Figs.\ \ref{fig:sec3fig4} and 
\ref{fig:sec3fig5}, respectively.  

The differences between the various potentials are larger  in calculating the
energy of nuclear matter. This is mainly due to the importance of the $^3S_1 -
^3D_1$ contribution as we discussed above. This is of course in line with
previous investigations, which showed that the predicted binding energy of
nuclear matter is correlated to the strength of the tensor force, expressed in
terms of the D-state probability obtained for the deuteron (see e.g.\
\cite{hko95}). This importance of the strength of the tensor force is also 
seen in the calculation of the binding energy of the triton in Ref.\ 
\cite{cdbonn,nhkg97}. 
The CD-Bonn potential yields a binding energy of $8.00$ MeV, 
the Nijm-I potential gives $7.72$ MeV while the Nijm-II 
yields $7.62$ MeV, the same as does the new Argonne potential \cite{v18}. 
  

\begin{figure}[htbp]
\setlength{\unitlength}{0.1bp}
\special{!
/gnudict 40 dict def
gnudict begin
/Color false def
/Solid false def
/gnulinewidth 5.000 def
/vshift -33 def
/dl {10 mul} def
/hpt 31.5 def
/vpt 31.5 def
/M {moveto} bind def
/L {lineto} bind def
/R {rmoveto} bind def
/V {rlineto} bind def
/vpt2 vpt 2 mul def
/hpt2 hpt 2 mul def
/Lshow { currentpoint stroke M
  0 vshift R show } def
/Rshow { currentpoint stroke M
  dup stringwidth pop neg vshift R show } def
/Cshow { currentpoint stroke M
  dup stringwidth pop -2 div vshift R show } def
/DL { Color {setrgbcolor Solid {pop []} if 0 setdash }
 {pop pop pop Solid {pop []} if 0 setdash} ifelse } def
/BL { stroke gnulinewidth 2 mul setlinewidth } def
/AL { stroke gnulinewidth 2 div setlinewidth } def
/PL { stroke gnulinewidth setlinewidth } def
/LTb { BL [] 0 0 0 DL } def
/LTa { AL [1 dl 2 dl] 0 setdash 0 0 0 setrgbcolor } def
/LT0 { PL [] 0 1 0 DL } def
/LT1 { PL [4 dl 2 dl] 0 0 1 DL } def
/LT2 { PL [2 dl 3 dl] 1 0 0 DL } def
/LT3 { PL [1 dl 1.5 dl] 1 0 1 DL } def
/LT4 { PL [5 dl 2 dl 1 dl 2 dl] 0 1 1 DL } def
/LT5 { PL [4 dl 3 dl 1 dl 3 dl] 1 1 0 DL } def
/LT6 { PL [2 dl 2 dl 2 dl 4 dl] 0 0 0 DL } def
/LT7 { PL [2 dl 2 dl 2 dl 2 dl 2 dl 4 dl] 1 0.3 0 DL } def
/LT8 { PL [2 dl 2 dl 2 dl 2 dl 2 dl 2 dl 2 dl 4 dl] 0.5 0.5 0.5 DL } def
/P { stroke [] 0 setdash
  currentlinewidth 2 div sub M
  0 currentlinewidth V stroke } def
/D { stroke [] 0 setdash 2 copy vpt add M
  hpt neg vpt neg V hpt vpt neg V
  hpt vpt V hpt neg vpt V closepath stroke
  P } def
/A { stroke [] 0 setdash vpt sub M 0 vpt2 V
  currentpoint stroke M
  hpt neg vpt neg R hpt2 0 V stroke
  } def
/B { stroke [] 0 setdash 2 copy exch hpt sub exch vpt add M
  0 vpt2 neg V hpt2 0 V 0 vpt2 V
  hpt2 neg 0 V closepath stroke
  P } def
/C { stroke [] 0 setdash exch hpt sub exch vpt add M
  hpt2 vpt2 neg V currentpoint stroke M
  hpt2 neg 0 R hpt2 vpt2 V stroke } def
/T { stroke [] 0 setdash 2 copy vpt 1.12 mul add M
  hpt neg vpt -1.62 mul V
  hpt 2 mul 0 V
  hpt neg vpt 1.62 mul V closepath stroke
  P  } def
/S { 2 copy A C} def
end
}
\begin{picture}(3600,2160)(0,0)
\special{"
gnudict begin
gsave
50 50 translate
0.100 0.100 scale
0 setgray
/Helvetica findfont 100 scalefont setfont
newpath
-500.000000 -500.000000 translate
LTa
600 251 M
2817 0 V
LTb
600 251 M
63 0 V
2754 0 R
-63 0 V
600 623 M
63 0 V
2754 0 R
-63 0 V
600 994 M
63 0 V
2754 0 R
-63 0 V
600 1366 M
63 0 V
2754 0 R
-63 0 V
600 1737 M
63 0 V
2754 0 R
-63 0 V
600 2109 M
63 0 V
2754 0 R
-63 0 V
600 251 M
0 63 V
0 1795 R
0 -63 V
913 251 M
0 63 V
0 1795 R
0 -63 V
1226 251 M
0 63 V
0 1795 R
0 -63 V
1539 251 M
0 63 V
0 1795 R
0 -63 V
1852 251 M
0 63 V
0 1795 R
0 -63 V
2165 251 M
0 63 V
0 1795 R
0 -63 V
2478 251 M
0 63 V
0 1795 R
0 -63 V
2791 251 M
0 63 V
0 1795 R
0 -63 V
3104 251 M
0 63 V
0 1795 R
0 -63 V
3417 251 M
0 63 V
0 1795 R
0 -63 V
600 251 M
2817 0 V
0 1858 V
-2817 0 V
600 251 L
LT0
2000 1946 M
180 0 V
600 669 M
12 8 V
46 29 V
47 31 V
46 31 V
47 29 V
46 27 V
46 27 V
47 28 V
46 27 V
46 26 V
47 24 V
46 24 V
47 24 V
46 24 V
47 25 V
46 25 V
46 23 V
47 22 V
46 22 V
47 23 V
46 23 V
47 25 V
46 25 V
46 25 V
47 26 V
46 25 V
47 25 V
46 24 V
46 24 V
47 24 V
46 22 V
47 20 V
46 21 V
46 19 V
47 20 V
46 19 V
47 21 V
46 22 V
47 21 V
46 22 V
46 22 V
47 21 V
46 22 V
47 22 V
46 21 V
47 21 V
46 20 V
46 20 V
47 20 V
46 18 V
46 16 V
47 16 V
46 16 V
47 16 V
46 16 V
46 16 V
47 16 V
46 19 V
47 19 V
46 19 V
47 20 V
20 9 V
LT1
2000 1846 M
180 0 V
600 656 M
12 7 V
46 27 V
47 29 V
46 29 V
47 26 V
46 24 V
46 25 V
47 25 V
46 24 V
46 22 V
47 20 V
46 20 V
47 20 V
46 21 V
47 20 V
46 21 V
46 19 V
47 18 V
46 19 V
47 18 V
46 19 V
47 21 V
46 21 V
46 22 V
47 21 V
46 21 V
47 21 V
46 20 V
46 20 V
47 19 V
46 18 V
47 16 V
46 15 V
46 16 V
47 15 V
46 15 V
47 17 V
46 17 V
47 18 V
46 18 V
46 18 V
47 18 V
46 18 V
47 17 V
46 18 V
47 17 V
46 16 V
46 16 V
47 15 V
46 14 V
46 11 V
47 12 V
46 11 V
47 10 V
46 11 V
46 11 V
47 12 V
46 14 V
47 15 V
46 15 V
47 16 V
20 7 V
LT2
2000 1746 M
180 0 V
600 642 M
12 7 V
46 25 V
47 28 V
46 28 V
47 24 V
46 23 V
46 23 V
47 24 V
46 22 V
46 21 V
47 18 V
46 18 V
47 17 V
46 18 V
47 18 V
46 19 V
46 18 V
47 18 V
46 18 V
47 18 V
46 19 V
47 20 V
46 19 V
46 20 V
47 19 V
46 19 V
47 19 V
46 18 V
46 18 V
47 17 V
46 16 V
47 13 V
46 13 V
46 13 V
47 12 V
46 13 V
47 14 V
46 15 V
47 15 V
46 15 V
46 15 V
47 15 V
46 15 V
47 3 V
46 26 V
47 15 V
46 13 V
46 14 V
47 13 V
46 11 V
46 10 V
47 10 V
46 9 V
47 9 V
46 9 V
46 10 V
47 9 V
46 11 V
47 11 V
46 12 V
47 12 V
20 5 V
LT3
2000 1646 M
180 0 V
600 644 M
12 7 V
46 26 V
47 28 V
46 28 V
47 25 V
46 23 V
46 23 V
47 24 V
46 23 V
46 21 V
47 20 V
46 18 V
47 18 V
46 19 V
47 19 V
46 19 V
46 19 V
47 18 V
46 19 V
47 19 V
46 20 V
47 21 V
46 20 V
46 21 V
47 20 V
46 21 V
47 20 V
46 19 V
46 19 V
47 18 V
46 17 V
47 14 V
46 15 V
46 13 V
47 14 V
46 13 V
47 16 V
46 16 V
47 16 V
46 17 V
46 16 V
47 17 V
46 16 V
47 16 V
46 16 V
47 15 V
46 15 V
46 14 V
47 14 V
46 13 V
46 11 V
47 11 V
46 11 V
47 10 V
46 10 V
46 11 V
47 11 V
46 13 V
47 13 V
46 13 V
47 14 V
20 6 V
LT4
2000 1546 M
180 0 V
600 645 M
12 8 V
46 28 V
47 26 V
46 26 V
47 25 V
46 24 V
46 25 V
47 25 V
46 24 V
46 23 V
47 21 V
46 20 V
47 18 V
46 18 V
47 18 V
46 17 V
46 21 V
47 22 V
46 22 V
47 23 V
46 22 V
47 20 V
46 19 V
46 20 V
47 20 V
46 21 V
47 23 V
46 23 V
46 23 V
47 23 V
46 24 V
47 25 V
46 24 V
46 24 V
47 23 V
46 24 V
47 21 V
46 21 V
47 20 V
46 19 V
46 19 V
47 19 V
46 18 V
47 17 V
46 17 V
47 17 V
46 17 V
46 17 V
47 18 V
46 17 V
46 19 V
47 19 V
46 19 V
47 20 V
46 20 V
46 20 V
47 21 V
46 21 V
47 22 V
46 23 V
47 22 V
20 11 V
stroke
grestore
end
showpage
}
\put(1944,1546){\makebox(0,0)[r]{${\cal S}$ with Argonne $V_{18}$}}
\put(1944,1646){\makebox(0,0)[r]{${\cal S}$ with Reid93}}
\put(1944,1746){\makebox(0,0)[r]{${\cal S}$ with Nijm II}}
\put(1944,1846){\makebox(0,0)[r]{${\cal S}$ with Nijm I}}
\put(1944,1946){\makebox(0,0)[r]{${\cal S}$ with CD-Bonn}}
\put(2008,21){\makebox(0,0){$n$ (fm$^{-3}$)}}
\put(100,1180){%
\special{ps: gsave currentpoint currentpoint translate
270 rotate neg exch neg exch translate}%
\makebox(0,0)[b]{\shortstack{${\cal S}(n)$ (MeV)}}%
\special{ps: currentpoint grestore moveto}%
}
\put(3417,151){\makebox(0,0){1}}
\put(3104,151){\makebox(0,0){0.9}}
\put(2791,151){\makebox(0,0){0.8}}
\put(2478,151){\makebox(0,0){0.7}}
\put(2165,151){\makebox(0,0){0.6}}
\put(1852,151){\makebox(0,0){0.5}}
\put(1539,151){\makebox(0,0){0.4}}
\put(1226,151){\makebox(0,0){0.3}}
\put(913,151){\makebox(0,0){0.2}}
\put(600,151){\makebox(0,0){0.1}}
\put(540,2109){\makebox(0,0)[r]{100}}
\put(540,1737){\makebox(0,0)[r]{80}}
\put(540,1366){\makebox(0,0)[r]{60}}
\put(540,994){\makebox(0,0)[r]{40}}
\put(540,623){\makebox(0,0)[r]{20}}
\put(540,251){\makebox(0,0)[r]{0}}
\end{picture}
     \caption{Symmetry energy ${\cal S}$ as function of density $n$.}
     \label{fig:sec3fig6}
\end{figure}
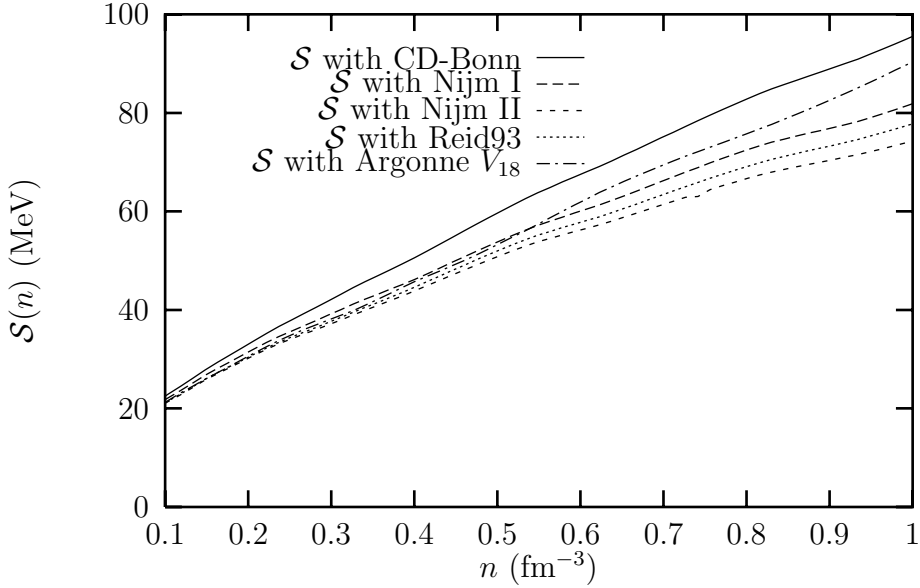

In Fig.\ \ref{fig:sec3fig5} we have plotted the results for 
pure neutron matter. 
For the density of  
$0.6$ fm$^{-3}$, we get
a contribution to the potential energy
of $-29.0$, $-30.2$, $-28.7$, $-28.1$ and $-24.0$ MeV
for the CD-Bonn, Nijm I, Nijm II, Reid93 and Argonne potentials, 
respectively, i.e., a difference of the order of few percents
for the first four potentials. The Argonne  $V_{18}$  potential \cite{v18}
deviates from the other four potentials due to slightly more repulsive 
contributions from $P$-waves, especially from the $^3P_1$ wave.
Similar results were also obtained recently in Ref.\ \cite{ap97}
for the new Argonne potential.
In the $T=1$ channel there are no contributions
from the $^3S_1$ channel and the tensor force plays therefore
a less important role. The differences in the $^1S_0$ channel
follow the discussion above in connection with Fig.\ \ref{fig:sec3fig1}.
Further 
differences are mainly due to contributions from the 
$^3P_1$ partial wave.

From the potential energy 
${\cal U}$ one can evaluate the total energy per nucleon
${\cal E}$ according to Eq.\ (\ref{eq:g_bhf}) for both 
symmetric nuclear matter
and pure neutron matter and identify the difference with the symmetry energy 
\begin{equation}
{\cal S} (n) = {\cal E} (n,\chi_p=0) - {\cal E} (n,\chi_p=1/2 )
\end{equation}
Results for these symmetry energies as a function of density $n$ are presented
in Fig.\ \ref{fig:sec3fig6}. None of the new potentials
predict a saturation of the symmetry energy as a function of density as was
the case for older local potentials discussed 
in Fig.\ \ref{fig:sec1fig1}. The
differences between the various predictions are smaller than those between the
older versions of the NN potential. The remaining discrepancies can be traced
back to the non-locality  of the potential considered, as we discussed above.
For the Argonne potential, where the results in symmetric nuclear matter are
almost identical to those from the Reid93 and the Nij II potentials
(see  e.g., Fig.\ \ref{fig:sec2fig2}) the increase is mainly due to the 
abovementioned more repulsive 
contributions from $P$-waves in neutron matter.

\begin{figure}[htbp]
\setlength{\unitlength}{0.1bp}
\special{!
/gnudict 40 dict def
gnudict begin
/Color false def
/Solid false def
/gnulinewidth 5.000 def
/vshift -33 def
/dl {10 mul} def
/hpt 31.5 def
/vpt 31.5 def
/M {moveto} bind def
/L {lineto} bind def
/R {rmoveto} bind def
/V {rlineto} bind def
/vpt2 vpt 2 mul def
/hpt2 hpt 2 mul def
/Lshow { currentpoint stroke M
  0 vshift R show } def
/Rshow { currentpoint stroke M
  dup stringwidth pop neg vshift R show } def
/Cshow { currentpoint stroke M
  dup stringwidth pop -2 div vshift R show } def
/DL { Color {setrgbcolor Solid {pop []} if 0 setdash }
 {pop pop pop Solid {pop []} if 0 setdash} ifelse } def
/BL { stroke gnulinewidth 2 mul setlinewidth } def
/AL { stroke gnulinewidth 2 div setlinewidth } def
/PL { stroke gnulinewidth setlinewidth } def
/LTb { BL [] 0 0 0 DL } def
/LTa { AL [1 dl 2 dl] 0 setdash 0 0 0 setrgbcolor } def
/LT0 { PL [] 0 1 0 DL } def
/LT1 { PL [4 dl 2 dl] 0 0 1 DL } def
/LT2 { PL [2 dl 3 dl] 1 0 0 DL } def
/LT3 { PL [1 dl 1.5 dl] 1 0 1 DL } def
/LT4 { PL [5 dl 2 dl 1 dl 2 dl] 0 1 1 DL } def
/LT5 { PL [4 dl 3 dl 1 dl 3 dl] 1 1 0 DL } def
/LT6 { PL [2 dl 2 dl 2 dl 4 dl] 0 0 0 DL } def
/LT7 { PL [2 dl 2 dl 2 dl 2 dl 2 dl 4 dl] 1 0.3 0 DL } def
/LT8 { PL [2 dl 2 dl 2 dl 2 dl 2 dl 2 dl 2 dl 4 dl] 0.5 0.5 0.5 DL } def
/P { stroke [] 0 setdash
  currentlinewidth 2 div sub M
  0 currentlinewidth V stroke } def
/D { stroke [] 0 setdash 2 copy vpt add M
  hpt neg vpt neg V hpt vpt neg V
  hpt vpt V hpt neg vpt V closepath stroke
  P } def
/A { stroke [] 0 setdash vpt sub M 0 vpt2 V
  currentpoint stroke M
  hpt neg vpt neg R hpt2 0 V stroke
  } def
/B { stroke [] 0 setdash 2 copy exch hpt sub exch vpt add M
  0 vpt2 neg V hpt2 0 V 0 vpt2 V
  hpt2 neg 0 V closepath stroke
  P } def
/C { stroke [] 0 setdash exch hpt sub exch vpt add M
  hpt2 vpt2 neg V currentpoint stroke M
  hpt2 neg 0 R hpt2 vpt2 V stroke } def
/T { stroke [] 0 setdash 2 copy vpt 1.12 mul add M
  hpt neg vpt -1.62 mul V
  hpt 2 mul 0 V
  hpt neg vpt 1.62 mul V closepath stroke
  P  } def
/S { 2 copy A C} def
end
}
\begin{picture}(3600,2160)(0,0)
\special{"
gnudict begin
gsave
50 50 translate
0.100 0.100 scale
0 setgray
/Helvetica findfont 100 scalefont setfont
newpath
-500.000000 -500.000000 translate
LTa
600 251 M
2817 0 V
LTb
600 251 M
63 0 V
2754 0 R
-63 0 V
600 716 M
63 0 V
2754 0 R
-63 0 V
600 1180 M
63 0 V
2754 0 R
-63 0 V
600 1645 M
63 0 V
2754 0 R
-63 0 V
600 2109 M
63 0 V
2754 0 R
-63 0 V
882 251 M
0 63 V
0 1795 R
0 -63 V
1445 251 M
0 63 V
0 1795 R
0 -63 V
2009 251 M
0 63 V
0 1795 R
0 -63 V
2572 251 M
0 63 V
0 1795 R
0 -63 V
3135 251 M
0 63 V
0 1795 R
0 -63 V
600 251 M
2817 0 V
0 1858 V
-2817 0 V
600 251 L
LT0
2114 1946 M
180 0 V
600 497 M
70 27 V
71 34 V
70 36 V
71 49 V
70 51 V
71 45 V
70 42 V
70 45 V
71 43 V
70 41 V
71 38 V
70 39 V
71 40 V
70 39 V
70 40 V
71 40 V
70 38 V
71 35 V
70 31 V
70 29 V
71 28 V
70 27 V
71 27 V
70 26 V
71 25 V
70 26 V
70 25 V
71 24 V
70 22 V
71 20 V
70 18 V
71 18 V
70 18 V
70 19 V
71 19 V
70 20 V
71 19 V
70 21 V
71 21 V
70 23 V
LT1
2114 1846 M
180 0 V
600 478 M
70 22 V
71 26 V
70 27 V
71 35 V
70 41 V
71 34 V
70 31 V
70 34 V
71 32 V
70 31 V
71 28 V
70 29 V
71 31 V
70 31 V
70 32 V
71 31 V
70 30 V
71 27 V
70 23 V
70 22 V
71 21 V
70 23 V
71 22 V
70 22 V
71 21 V
70 21 V
70 19 V
71 19 V
70 17 V
71 14 V
70 14 V
71 13 V
70 13 V
70 15 V
71 15 V
70 16 V
71 16 V
70 18 V
71 20 V
70 21 V
LT2
2114 1746 M
180 0 V
600 458 M
70 20 V
71 23 V
70 22 V
71 26 V
70 34 V
71 29 V
70 26 V
70 29 V
71 28 V
70 28 V
71 26 V
70 27 V
71 27 V
70 28 V
70 28 V
71 28 V
70 26 V
71 23 V
70 19 V
70 17 V
71 16 V
70 17 V
71 16 V
70 16 V
71 17 V
70 18 V
70 18 V
71 17 V
70 16 V
71 11 V
70 11 V
71 9 V
70 9 V
70 10 V
71 10 V
70 10 V
71 11 V
70 11 V
71 15 V
70 15 V
LT3
2114 1646 M
180 0 V
600 461 M
70 20 V
71 25 V
70 23 V
71 28 V
70 36 V
71 31 V
70 28 V
70 30 V
71 31 V
70 29 V
71 28 V
70 29 V
71 30 V
70 30 V
70 31 V
71 30 V
70 29 V
71 26 V
70 21 V
70 20 V
71 18 V
70 19 V
71 18 V
70 18 V
71 19 V
70 19 V
70 19 V
71 18 V
70 17 V
71 14 V
70 12 V
71 13 V
70 12 V
70 12 V
71 13 V
70 14 V
71 14 V
70 14 V
71 16 V
70 16 V
LT4
2114 1546 M
180 0 V
600 463 M
70 24 V
71 19 V
70 22 V
71 31 V
70 42 V
71 38 V
70 32 V
70 29 V
71 28 V
70 31 V
71 33 V
70 32 V
71 31 V
70 31 V
70 35 V
71 38 V
70 40 V
71 41 V
70 41 V
70 40 V
71 38 V
70 33 V
71 31 V
70 29 V
71 26 V
70 25 V
70 24 V
71 23 V
70 21 V
71 21 V
70 21 V
71 20 V
70 21 V
70 19 V
71 20 V
70 20 V
71 23 V
70 25 V
71 28 V
70 32 V
stroke
grestore
end
showpage
}
\put(2054,1546){\makebox(0,0)[r]{$\chi_p$ with Argonne $V_{18}$}}
\put(2054,1646){\makebox(0,0)[r]{$\chi_p$ with Reid93}}
\put(2054,1746){\makebox(0,0)[r]{$\chi_p$ with Nijm II}}
\put(2054,1846){\makebox(0,0)[r]{$\chi_p$ with Nijm I}}
\put(2054,1946){\makebox(0,0)[r]{$\chi_p$ with CD-Bonn}}
\put(2008,21){\makebox(0,0){$n$ (fm$^{-3}$)}}
\put(100,1180){%
\special{ps: gsave currentpoint currentpoint translate
270 rotate neg exch neg exch translate}%
\makebox(0,0)[b]{\shortstack{$\chi_p$}}%
\special{ps: currentpoint grestore moveto}%
}
\put(3135,151){\makebox(0,0){1}}
\put(2572,151){\makebox(0,0){0.8}}
\put(2009,151){\makebox(0,0){0.6}}
\put(1445,151){\makebox(0,0){0.4}}
\put(882,151){\makebox(0,0){0.2}}
\put(540,2109){\makebox(0,0)[r]{0.2}}
\put(540,1645){\makebox(0,0)[r]{0.15}}
\put(540,1180){\makebox(0,0)[r]{0.1}}
\put(540,716){\makebox(0,0)[r]{0.05}}
\put(540,251){\makebox(0,0)[r]{0}}
\end{picture}
     \caption{Proton fraction $\chi_p$ for $\beta$-stable 
              matter as function of density $n$}
     \label{fig:sec3fig7}
\end{figure}
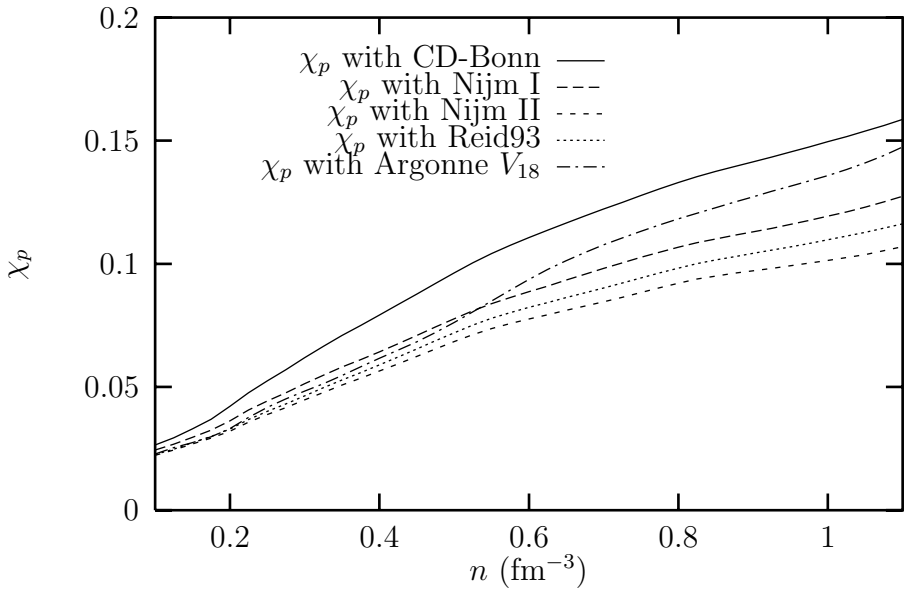

From the differences in symmetry energies
one would expect that properties like proton fractions in
$\beta$-stable matter will be influenced. This in turn has consequences for 
the cooling history of a neutron star. 
Thus, for the sake of completeness, we show 
in Fig.\ \ref{fig:sec3fig7} resulting
proton fractions in $\beta$-stable matter. 
The proton fraction in $\beta$-equilibrium is
determined by imposing the relevant equilibrium conditions on the processes
$e^{-}+p \rightarrow n+\nu_{e}$
and $e^{-}\rightarrow \mu^{-}+\overline{\nu}_{\mu}+\nu_{e}$.
The conditions for $\beta$-equilibrium require that 
$\mu_{n}=\mu_{p}+\mu_{e}$,
where $\mu_{i}$ is the chemical potential of particle species $i$ ,
and that 
charge is conserved 
$n_{p}=n_{e}$.
We include muons, which appear at densities close to the saturation
density for nuclear matter. 
We assume that the neutrinos do 
not contribute.
The proton and neutron chemical potentials are determined  from 
the energy per particle ${\cal E}(n,\chi_p)$ where we calculate 
the latter quantity for several proton fractions $\chi_p$ and 
impose the above equilibrium conditions for $\beta$-stable matter.

It is seen from Fig.\ \ref{fig:sec3fig7} that the differences
reflected in the symmetry energy appear also for the proton fractions.
For the CD-Bonn potential the direct URCA process can occur at
a density of $0.88$ fm$^{-3}$, for the Nijm I it starts
at $1.25$ fm$^{-3}$ while for the Reid93 potential one reaches
the critical density at $1.36$ fm$^{-3}$. The Argonne potential
allows for the direct URCA process at a density of  $1.05$ fm$^{-3}$.
For the Nijm II we were not able to get the direct URCA process
for densities below $1.5$ fm$^{-3}$ MeV.

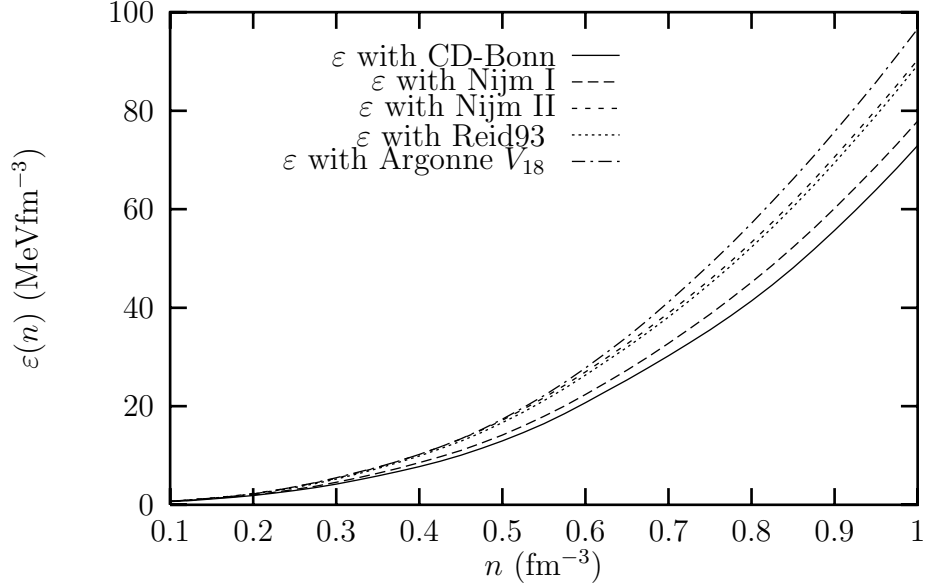
\begin{figure}[htbp]
\setlength{\unitlength}{0.1bp}
\special{!
/gnudict 40 dict def
gnudict begin
/Color false def
/Solid false def
/gnulinewidth 5.000 def
/vshift -33 def
/dl {10 mul} def
/hpt 31.5 def
/vpt 31.5 def
/M {moveto} bind def
/L {lineto} bind def
/R {rmoveto} bind def
/V {rlineto} bind def
/vpt2 vpt 2 mul def
/hpt2 hpt 2 mul def
/Lshow { currentpoint stroke M
  0 vshift R show } def
/Rshow { currentpoint stroke M
  dup stringwidth pop neg vshift R show } def
/Cshow { currentpoint stroke M
  dup stringwidth pop -2 div vshift R show } def
/DL { Color {setrgbcolor Solid {pop []} if 0 setdash }
 {pop pop pop Solid {pop []} if 0 setdash} ifelse } def
/BL { stroke gnulinewidth 2 mul setlinewidth } def
/AL { stroke gnulinewidth 2 div setlinewidth } def
/PL { stroke gnulinewidth setlinewidth } def
/LTb { BL [] 0 0 0 DL } def
/LTa { AL [1 dl 2 dl] 0 setdash 0 0 0 setrgbcolor } def
/LT0 { PL [] 0 1 0 DL } def
/LT1 { PL [4 dl 2 dl] 0 0 1 DL } def
/LT2 { PL [2 dl 3 dl] 1 0 0 DL } def
/LT3 { PL [1 dl 1.5 dl] 1 0 1 DL } def
/LT4 { PL [5 dl 2 dl 1 dl 2 dl] 0 1 1 DL } def
/LT5 { PL [4 dl 3 dl 1 dl 3 dl] 1 1 0 DL } def
/LT6 { PL [2 dl 2 dl 2 dl 4 dl] 0 0 0 DL } def
/LT7 { PL [2 dl 2 dl 2 dl 2 dl 2 dl 4 dl] 1 0.3 0 DL } def
/LT8 { PL [2 dl 2 dl 2 dl 2 dl 2 dl 2 dl 2 dl 4 dl] 0.5 0.5 0.5 DL } def
/P { stroke [] 0 setdash
  currentlinewidth 2 div sub M
  0 currentlinewidth V stroke } def
/D { stroke [] 0 setdash 2 copy vpt add M
  hpt neg vpt neg V hpt vpt neg V
  hpt vpt V hpt neg vpt V closepath stroke
  P } def
/A { stroke [] 0 setdash vpt sub M 0 vpt2 V
  currentpoint stroke M
  hpt neg vpt neg R hpt2 0 V stroke
  } def
/B { stroke [] 0 setdash 2 copy exch hpt sub exch vpt add M
  0 vpt2 neg V hpt2 0 V 0 vpt2 V
  hpt2 neg 0 V closepath stroke
  P } def
/C { stroke [] 0 setdash exch hpt sub exch vpt add M
  hpt2 vpt2 neg V currentpoint stroke M
  hpt2 neg 0 R hpt2 vpt2 V stroke } def
/T { stroke [] 0 setdash 2 copy vpt 1.12 mul add M
  hpt neg vpt -1.62 mul V
  hpt 2 mul 0 V
  hpt neg vpt 1.62 mul V closepath stroke
  P  } def
/S { 2 copy A C} def
end
}
\begin{picture}(3600,2160)(0,0)
\special{"
gnudict begin
gsave
50 50 translate
0.100 0.100 scale
0 setgray
/Helvetica findfont 100 scalefont setfont
newpath
-500.000000 -500.000000 translate
LTa
600 251 M
2817 0 V
LTb
600 251 M
63 0 V
2754 0 R
-63 0 V
600 623 M
63 0 V
2754 0 R
-63 0 V
600 994 M
63 0 V
2754 0 R
-63 0 V
600 1366 M
63 0 V
2754 0 R
-63 0 V
600 1737 M
63 0 V
2754 0 R
-63 0 V
600 2109 M
63 0 V
2754 0 R
-63 0 V
600 251 M
0 63 V
0 1795 R
0 -63 V
913 251 M
0 63 V
0 1795 R
0 -63 V
1226 251 M
0 63 V
0 1795 R
0 -63 V
1539 251 M
0 63 V
0 1795 R
0 -63 V
1852 251 M
0 63 V
0 1795 R
0 -63 V
2165 251 M
0 63 V
0 1795 R
0 -63 V
2478 251 M
0 63 V
0 1795 R
0 -63 V
2791 251 M
0 63 V
0 1795 R
0 -63 V
3104 251 M
0 63 V
0 1795 R
0 -63 V
3417 251 M
0 63 V
0 1795 R
0 -63 V
600 251 M
2817 0 V
0 1858 V
-2817 0 V
600 251 L
LT0
2114 1946 M
180 0 V
600 263 M
78 4 V
79 6 V
78 6 V
78 7 V
78 9 V
79 10 V
78 11 V
78 13 V
78 15 V
79 16 V
78 17 V
78 18 V
78 20 V
79 23 V
78 26 V
78 28 V
78 31 V
79 34 V
78 38 V
78 41 V
78 43 V
79 43 V
78 45 V
78 46 V
78 48 V
79 50 V
78 53 V
78 56 V
78 60 V
79 65 V
78 69 V
78 72 V
78 75 V
79 79 V
78 81 V
78 85 V
LT1
2114 1846 M
180 0 V
600 264 M
78 4 V
79 7 V
78 6 V
78 7 V
78 10 V
79 11 V
78 12 V
78 15 V
78 16 V
79 18 V
78 19 V
78 21 V
78 22 V
79 24 V
78 28 V
78 30 V
78 34 V
79 37 V
78 39 V
78 43 V
78 45 V
79 47 V
78 49 V
78 52 V
78 54 V
79 56 V
78 59 V
78 61 V
78 64 V
79 69 V
78 71 V
78 75 V
78 78 V
79 81 V
78 83 V
78 87 V
LT2
2114 1746 M
180 0 V
600 264 M
78 6 V
79 7 V
78 7 V
78 10 V
78 11 V
79 14 V
78 15 V
78 17 V
78 20 V
79 21 V
78 24 V
78 25 V
78 28 V
79 30 V
78 34 V
78 37 V
78 41 V
79 44 V
78 47 V
78 50 V
78 52 V
79 54 V
78 57 V
78 59 V
78 62 V
79 65 V
78 68 V
78 72 V
78 75 V
79 79 V
78 82 V
78 86 V
78 88 V
79 91 V
78 92 V
78 95 V
LT3
2114 1646 M
180 0 V
600 264 M
78 6 V
79 6 V
78 7 V
78 9 V
78 11 V
79 13 V
78 15 V
78 17 V
78 19 V
79 21 V
78 23 V
78 24 V
78 27 V
79 29 V
78 33 V
78 37 V
78 39 V
79 44 V
78 46 V
78 50 V
78 52 V
79 53 V
78 56 V
78 58 V
78 61 V
79 65 V
78 67 V
78 71 V
78 73 V
79 79 V
78 81 V
78 85 V
78 88 V
79 91 V
78 93 V
78 96 V
LT4
2114 1546 M
180 0 V
600 264 M
78 6 V
79 6 V
78 7 V
78 9 V
78 12 V
79 15 V
78 16 V
78 18 V
78 20 V
79 20 V
78 23 V
78 25 V
78 28 V
79 31 V
78 35 V
78 39 V
78 43 V
79 46 V
78 50 V
78 54 V
78 57 V
79 60 V
78 64 V
78 68 V
78 70 V
79 73 V
78 75 V
78 79 V
78 82 V
79 85 V
78 87 V
78 91 V
78 92 V
79 96 V
78 99 V
78 101 V
stroke
grestore
end
showpage
}
\put(2054,1546){\makebox(0,0)[r]{${\varepsilon}$ with Argonne $V_{18}$ }}
\put(2054,1646){\makebox(0,0)[r]{${\varepsilon}$ with Reid93 }}
\put(2054,1746){\makebox(0,0)[r]{${\varepsilon}$ with Nijm II}}
\put(2054,1846){\makebox(0,0)[r]{${\varepsilon}$ with Nijm I}}
\put(2054,1946){\makebox(0,0)[r]{${\varepsilon}$ with CD-Bonn}}
\put(2008,21){\makebox(0,0){$n$ (fm$^{-3}$)}}
\put(100,1180){%
\special{ps: gsave currentpoint currentpoint translate
270 rotate neg exch neg exch translate}%
\makebox(0,0)[b]{\shortstack{${\varepsilon}(n)$ (MeVfm$^{-3}$)}}%
\special{ps: currentpoint grestore moveto}%
}
\put(3417,151){\makebox(0,0){1}}
\put(3104,151){\makebox(0,0){0.9}}
\put(2791,151){\makebox(0,0){0.8}}
\put(2478,151){\makebox(0,0){0.7}}
\put(2165,151){\makebox(0,0){0.6}}
\put(1852,151){\makebox(0,0){0.5}}
\put(1539,151){\makebox(0,0){0.4}}
\put(1226,151){\makebox(0,0){0.3}}
\put(913,151){\makebox(0,0){0.2}}
\put(600,151){\makebox(0,0){0.1}}
\put(540,2109){\makebox(0,0)[r]{100}}
\put(540,1737){\makebox(0,0)[r]{80}}
\put(540,1366){\makebox(0,0)[r]{60}}
\put(540,994){\makebox(0,0)[r]{40}}
\put(540,623){\makebox(0,0)[r]{20}}
\put(540,251){\makebox(0,0)[r]{0}}
\end{picture}
     \caption{Equation of state $\varepsilon$ for $\beta$-stable 
              matter as function of density $n$}
     \label{fig:sec3fig8}
\end{figure}

Finally, in Fig.\ \ref{fig:sec3fig8} we plot the final equation of state
$\varepsilon$ (which also includes the contribution from leptons)
for  $\beta$-stable matter. Since the proton fractions are not too large,
see Fig.\ \ref{fig:sec3fig7}, the $^3S_1$ contribution with $T_z=0$ plays 
a less significant role than that seen in Fig.\ \ref{fig:sec3fig2}. Thus,
the main contribution to the differences between the various
potentials arises from the $T_z=1$ channel.
The results for $\beta$-stable matter shown in Fig.\ \ref{fig:sec3fig8}
are gratifying in the sense that with the present high-quality potentials 
one does not have the large differences between 
potentials
as seen e.g.\ in Ref.\ \cite{wff88}. 
This means also that a calculation
like the present with two-body forces only and employing
new high-quality potentials will yield a mass-radius relationship for a 
neutron star which is rather similar for all potentials considered.

\section{Conclusions}\label{sec:sec4}

The most recent parametrizations of the nucleon-nucleon potential have been
employed to evaluate the energy of asymmetric nuclear matter at high densities
using the non-relativistic Brueckner-Hartree-Fock (BHF) approximation.
An attempt has been made to relate the differences in the predicted energies to
basic features of these interactions which are essentially phase-shift
equivalent.  

We find that the energies predicted for pure neutron matter and $\beta$-stable
matter are quite similar for all modern NN potentials. This may indicate that 
inconsistencies between various models for the NN potential in 
the isospin $T=1$ 
channel have been diminished by the improved fit to the data. Differences are
larger for the calculated energy of 
symmetric nuclear matter. All symmetry energies resulting from the different
potentials increase with density, no saturation is observed, as was the case 
for older models of the NN potential\cite{wsa84,reid68}. The remaining
discrepancies can be traced back to the inclusion of
non-local contributions in the short-range components of the NN interaction as
well as in the One-Pion-Exchange contribution. 
The non-localities included in
the CD-Bonn potential\cite{cdbonn} and partly 
also in the Nijm-I potential of
the Nijmegen group\cite{nim} lead to more binding energy in 
nuclear matter. These
non-localities are based on the meson-exchange model of the NN interaction.
Their inclusion may be considered as an improvement compared to
purely local potentials like Nijm-II, Reid93 and Argonne $V_{18}$\cite{v18}. 
Additional non-local contributions may arise from short-range 
quark-gluon exchange.

The calculations have been performed using a simple 
many-body scheme, a non-relativistic Brueckner-Hartree-Fock 
approach. The reason being that this method
allows one in a direct way to relate certain
many-body terms directly to the $T$-matrix.
More complicated many-body terms have to be included
in order to obtain a realistic EOS and symmetry energy. Three-body
forces \cite{wff88,ap97}
and/or relativistic effects \cite{ehobo96} are known to add 
repulsion at higher densities.
These effects may therefore yield a more repulsive EOS and stiffer
symmetry energy.

\ack{
This work was supported in part by The Research Council of Norway
(Programme for Supercomputing) through a grant for computing time
and by the U.S.\ National Science Foundation through Grant 
No.~PHY-9603097.
Many discussions with Marcello Baldo, Fiorella Burgio, 
\O ystein Elgar\o y, 
Eivind Osnes, Vijay Pandharipande, 
Steven Pieper, Angels Ramos and Robert Wiringa are greatly acknowledged. This
work has been completed at the European Centre for Theoretical Nuclear Physics
and Related Areas, ECT*, 
Trento, Italy. We acknowledge the hospitality of the ECT*.}


\begin{thebibliography}{200}
\bibitem{pr95} C.J.\ Pethick and D.G.\ Ravenhall, Annu.\ Rev.\ Nucl.\
Part.\ Phys.\ 45 (1995) 429. 
\bibitem{prakash94} M.\ Prakash, Phys.\ Rep.\ 242 (1994) 191.
\bibitem{bkr97} B.-A.\ Li, C.M.\ Ko and Z.\ Ren, Phys.\ Rev.\ Lett.\ 78 (1997) 1644.
\bibitem{tanihata95} I.\ Tanihata, Prog.\ Part.\ Nucl.\ Phys. 35 (1995) 505.
\bibitem{hjj95} P.G.\ Hansen, A.S.\ Jensen and B.\ Jonson, Annu.\ Rev.\
Nucl.\ Part.\ Sci.\ 45 (1995) 591.
\bibitem{fm79} B.L.\ Friman and O.V.\ Maxwell, 
Ap.\ J.\ 232 (1979) 541.
\bibitem{wff88} R.B.\ Wiringa, V.\ Fiks and A.\ Fabrocini, Phys.\ Rev.\
C 38 (1988) 1010.
\bibitem{wsa84} R.B.\ Wiringa, R.A.\ Smith and T.L.\
Ainsworth, Phys.\ Rev.\ C 29 (1984) 1207.
\bibitem{ehobo96} L.\ Engvik, E.\ Osnes, M.\ Hjorth-Jensen, G.\ Bao and
E.\ \O stgaard, Ap.\ J.\ 469  (1996) 794.
\bibitem{bbb96} M.\ Baldo, G.F.\ Burgio and I.\ Bombaci, submitted to
Phys.\ Rev.\ Lett.
\bibitem{reid68} R.V.\ Reid, Ann.\ Phys.\ 50  (1968) 411.
\bibitem{paris80} M.\ Lacombe {\em et al}, Phys.\ Rev.\ C {\bf 21}, 861 (1980).
\bibitem{cdbonn} R.\ Machleidt, F.\ Sammarruca and Y.\ Song,
Phys.\ Rev.\ C 53 (1996) R1483.
\bibitem{v18} R.B.\ Wiringa, V.G.J.\ Stoks and R.\ Schiavilla, 
Phys.\ Rev.\ C 51 (1995) 38.
\bibitem{nim} V.G.J.\ Stoks, R.A.\ M.\ Klomp, C.P.F.\ Terheggen 
and J.J.\
de Swart, Phys.\ Rev.\ C 49  (1994) 2950.
\bibitem{mac89} R.\ Machleidt, Adv.\ Nucl.\ Phys.\ 19 (1989) 185. 
\bibitem{nhkg97} A.\ Nogga, D.\ H\"uber, H.\ Kamada and W.\ Gl\"ockle, preprint nucl-th/9704001.
\bibitem{stoks97}  V.G.\ J.\ Stoks, private communication.
\bibitem{mahaux85} J.P.\ Jeukenne, A.\ Lejeune and C.\ Mahaux,
Phys.\ Rep.\  25 (1976) 83;
C.\ Mahaux, P.F.\ Bortignon, R.A.\ Broglia and C.H.\ Dasso,
 Phys.\ Rep.\ 120 (1985) 1.
\bibitem{trebal} M.\ Baldo, contribution to ``International Workshop on Microscopic
Many-Body Methods'', ECT*, Trento, Italy (May 26-june 6, 1997).
\bibitem{day81} B.D.\ Day, Phys.\ Rev.\ C 24 (1981) 1203.
\bibitem{ap97} A.\ Akmal and V.R.\ Pandharipande, preprint
nucl-th/9705013 and submitted to Phys.\ Rev.\ C.
\bibitem{hko95} M.\ Hjorth-Jensen, T.T.S.\ Kuo and E.\ Osnes,
Phys.\ Rep.\ 261 (1995) 125.

\end{thebibliography}
\end{document}